# Material exploration through active learning – METAL

Joakim Brorsson,[1] Henrik Klein Moberg,[1] Joel Hildingsson,[1] Jonatan Gastaldi,[2]
Tobias Mattisson,[2] and Anders Hellman[1, *]

[1]*Chalmers University of Technology, Department of Physics, SE-412 96 Gothenburg, Sweden*
[2]*Chalmers University of Technology, Department of Space, Earth and Environment, SE-412 96 Gothenburg, Sweden*

The discovery and design of new materials are paramount in the development of green technologies. High entropy oxides represent one such group that has only been tentatively explored, mainly due to the inherent problem of navigating vast compositional spaces. Thanks to the emergence of machine learning, however, suitable tools are now readily available. Here, the task of finding oxygen carriers for chemical looping processes has been tackled by leveraging active learning-based strategies combined with first-principles calculations. High efficiency and efficacy have, moreover, been achieved by exploiting the power of recently developed machine learning interatomic potentials. Firstly, the proposed approaches were validated based on an established computational framework for identifying high entropy perovskites that can be used in chemical looping air separation and dry reforming. Chief among the insights thus gained was the identification of the best performing strategies, in the form of greedy or Thompson-based sampling based on uncertainty estimates obtained from Gaussian processes. Building on this newfound knowledge, the concept was applied to a more complex problem, namely the discovery of high entropy oxygen carriers for chemical looping oxygen uncoupling. This resulted in both qualitative as well as quantitative outcomes, including lists of specific materials with high oxygen transfer capacities and configurational entropies. Specifically, the best candidates were based on the known oxygen carrier $CaMnO_3$ but also contained a variety of additional species, of which some, e.g., Ti; Co; Cu; and Ti, were expected while others were not, e.g., Y and Sm. The results suggest that adopting active learning approaches is critical in materials discovery, given that these methods are already shifting research practice and soon will be the norm.

## I. INTRODUCTION

Throughout human history, advances in materials discovery have powered technological innovation and shaped societal change. Obvious examples include the successive identification of better metals and alloys, which began in prehistoric times with copper, followed by bronze and then iron [1]. Skipping forward to the modern era, numerous alloys have now been produced that incorporate a wide variety of different elements and have been tailored to suit specific applications. Even so, progress is still being made, as is exemplified by the relatively recent identification of high entropy alloys (HEAs) as a novel class of materials that possess unique properties [2]. Later, this concept was extended to include other types of inorganic compounds, including oxides, carbides, and pnictides, which are together labelled as high entropy materials (HEMs) [3].

A fundamental issue related to the exploration of new materials is the vastness of the compositional, or configurational, space. This is especially true for HEMs, in which five or more elements typically occupy the same sublattice [4]. Due to the enormous number of variations, it is impossible to sample a significant fraction of them. As such, a conventional trial-and-error approach is almost guaranteed to fail to identify the best possible candidate for a given application. While existing knowledge and data can help guide the search, this would likely reduce the chance of discovering a completely novel material.

In recent years, machine learning (ML) has emerged as a potent tool for solving problems involving enormous amounts of data. Active learning (AL), in particular, is a technique that has only begun to be explored, especially in the context of material development. Yet, the few studies conducted so far have revealed the great potency of this approach, underscoring the likelihood that it will become more widespread, or even the new norm, in a not so distant future. Such strategies have, in fact, allowed the millions or even billions of candidate materials to be generated and tested, leading to the discovery of several hundred thousand stable compounds [5–7]. It is worth noting, however, that this type brute-force screening require enormous computational resources while still only probing a tiny fraction of the compositional space spanned by HEMs. In addition, the resulting a collection of materials is only known to fulfil a limited set of requirements, chiefly related to their thermodynamically stability at 0K. Consequently, some type high-throughput screening process would be required [8], together with additional, possibly demanding, calculations of the properties of interest, in order to identify material candidates for a specific application. For this reason, balanced and more streamlined approaches, such as the one originally presented by Rao et al. appear more promising [9]. Specifically, they introduced a procedure that utilizes AL to minimize amount of data required for training and used it to search for high entropy Invar alloys with high thermal stabilities and low thermal expansion coefficients, thereby discovering several previously untested materials.

Due to the limited use of AL strategies in the study of inorganic systems [9], the aim of this study was to develop an such an approach for discovering new, and potentially complex, materials, which are optimized for a particular application, based on sparse data. More precisely, this involves training a neural network (NN)—more precisely, a Wasserstein auto-encoder (WAE)—on compositions together with a target property, which could, in principle, be obtained by any method, such as experiments or first-principles calculations. In this work, we estimate the properties of interest using an machine learning interatomic potential (MLIP), which has been trained on first-principles data and, thus, represents a validated computational method that is both efficient and effective. Because of the urgent need for developing technologies that can help combat climate change, the proposed method is used to identify oxygen carriers (OCs) for chemical looping (CL) in order to demonstrate its versatility. Such a focus is further validated by the fact that complex materials, including high entropy oxides (HEOs), have shown promise for this type of process, which not only allows efficient fuel conversion with low emissions, but also represents a break-through concept for carbon capture [10]. In particular, three distinct types are considered: chemical looping dry reforming (CLDR), for CO and $H_2$ generation; chemical looping air separation (CLAS), for $O_2$ production; and chemical looping oxygen uncoupling (CLOU), for fuel combustion.

In the section that follows, an elaborate description of the computational method is provided, in terms of training the NN; sampling the latent space; and evaluating the thermodynamic properties. The results obtained when applying this approach to the problem of finding perovskite OCs are presented afterward, starting with CLDR and then CLAS followed by CLOU. Finally, the key advantages and future prospects of the proposed methodology are highlighted.

## II. METHOD

The AL-based strategy employed in this study has been inspired by the work of Rao et al. [9]. Specifically, it is based on the same concept for identifying promising materials for specific applications and is especially well suited for handling intricate compositional spaces, such as those as encountered when considering HEMs. Even so, the approaches differ in many ways, for instance in terms of the NN design, sampling procedure, and process for evaluating candidates. In addition, the goal has been to find OCs, in the form complex oxides, for CL, as this is a potential breakthrough energy conversion technology.

Before the AL cycle can be initiated, one must first generate an initial dataset, here in the form of a structure database (see section IIA). When the latter is in place, one can proceed with first step, which involves using the NN, in the form of a WAE, to compress composition data, as well as target properties, into a multi-dimensional latent space, represented by a Gaussian mixture model (GMM) (see section IIB). Thereafter, a suitable method is utilized for generating candidates; preferably one that balances exploration and exploitation (see section IIC). The structural and thermodynamic properties of the sampled compositions are determined using first principle calculations (see section IID and section IIA3). To be precise, the latter were performed with help of a MLIP (CHGNet) that has been trained on density functional theory (DFT) data and, thus, achieves almost the same level of accuracy at a fraction of the cost. As a final step, the training database is updated with the resulting dataset, thereby completing the AL cycle (see Figure 1).

### A. Step 1: Expand the database

#### *1. Structure generation*

To verify the proposed computational approach, the original training data was selected in the same way as in the study by Wang et al. [11]. In particular, this meant considering all 2401 possible configurations of a cubic 40 atom $Sr_{1-x}A_xFe_{1-y}B_yO_{3-\delta}$ supercell, where A = Ca, K, Y, Ba, La, Sm; B = Co, Cu, Mn, Mg, Ni, Ti; and $x, y \in \{0, 0.125, 0.25, 0.375, 0.5, 0.625, 0.75, 0.875, 1\}$. As is detailed in the next subsection, the property of interest, which here represents the probability that any of the vacancy formation energies are found within processspecific intervals, was evaluated for special quasi-random structures (SQSs) with $\delta \in \{0, 0.125, 0.25, 0.375, 0.5\}$. To account for a possible decomposition of the perovskite phase at $\delta$ = 0.5, the corresponding brownmillerite structure, $Sr_{2-2x}A_{2x}Fe_{2-2y}B_{2y}O_5$, was also included. Similarly to Wang et al. [11], such a transition was only allowed when considering applications below 800 °C, based on the argument that the oxygen vacancies should be randomly distributed above this temperature. In both cases, the most stable entries corresponding to the chemical formulas $SrFeO_3$ and $Sr_2Fe_2BO_5$ were retrieved from the Materials Project [12] using the pymatgen toolkit [13].

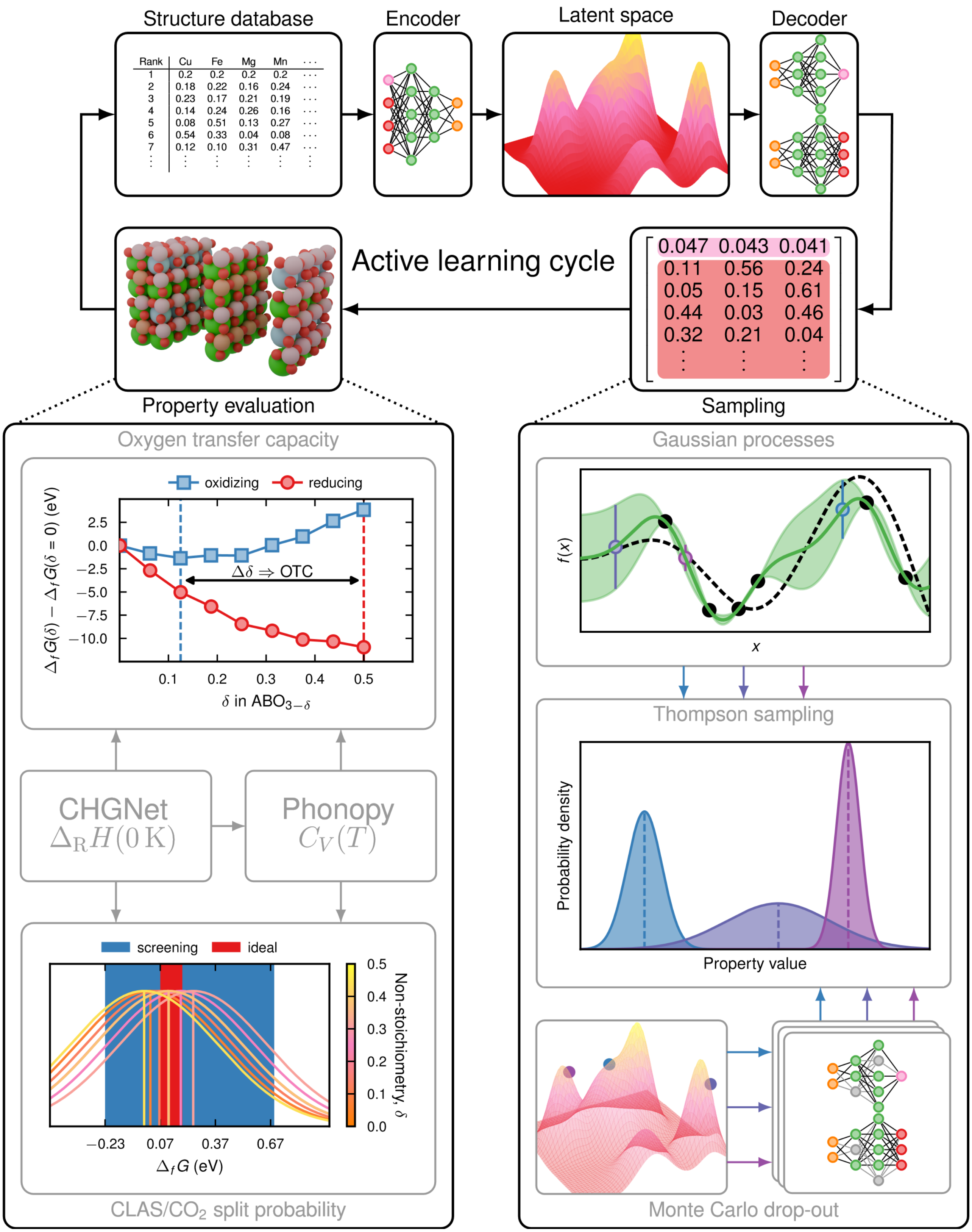

FIG. 1. Conceptual sketch of the main steps in a single AL cycle. First, a WAE is used to encode compositions and target properties, contained in a database, into a latent space, represented by a GMM. Next, promising candidates are selected using a chosen method, such as Greedy or Thompson sampling, based on property values and uncertainty estimates obtained either via GP or MC dropout. As a validation, either the OTC or the probability that the material is a suitable OC for a given type of CL process is, subsequently, calculated with the help of CHGNet. The loop is closed by feeding the resulting compositions, as well as the computed values on the target property, into the database.

When the same AL strategy was applied to the problem of identifying OCs for CLOU, larger supercells, with 80 atoms and up to seven distinct metallic elements, were generated. Specifically, this resulted in a training data set consisting of 1815 random $ABO_3$configurations. Since the structural symmetry of perovskite materials is known to vary with the temperature and vacancy concentration, both orthorhombic and cubic symmetries were considered, in addition to the brownmillerite variant [14]. The expansion of the unit cell also allowed intermediate vacancy concentrations to be included, namely $\delta \in \{0.0625, 0.1825, 0.3125, 0.4375\}$. Calculations were, thus, performed for 19 distinct SQSs for each composition, even though the estimates of the vacancy formation energies were based on the most stable structure for a given $\delta$.

### *2. Filtering*

To assess the synthesizability of the perovskite compositions, for the purpose of filtering out those that are likely to be unstable, the Bartel tolerance ($\tau$) [15] and charge balance ($\Delta q_e$) were computed using the same procedure as Wang et al.[11]. In the verification study, all structures in the original training data set $\tau > 4.3$ or $\Delta q_n = 0$ were filtered out, meaning that 2069 remained. For the samples produced by the neural network, the Goldschmidt, rather than Bartel, tolerance [16] was used as a measure of stability and, specifically, requiring it to be within the interval
$0.9 \leq t \leq 1.1$. In the context of finding OCs for CLOU, meanwhile, the limit originally proposed by Bartel et al., i.e. $\tau > 4.18$, was used, together with the charge neutrality requirement, for filtering the training data, which left 1066 compositions. Note that an additional constraint was placed on the samples generated during the AL cycles to ensure that the selected candidates satisfied the HEM criteria, namely that the configurational entropy must exceed the ideal gas constant (R) by a factor of at least 1.5 ($S_{\mathrm{conf}} \geq 1.5$R).

### *3. First principles*

For the sake of leveraging speed and accuracy, one of the recently developed[17–22] universal MLIPs was used, namely CHGNet, to calculate the vacancy formation energies. As is detailed by Deng et al. [17], this potential has been trained on forces, energies, tensors, and magnetic moments for about 1.6 million trajectories available from the Materials Project database [12]. In contrast to comparative models, e.g. M3GNet [18] and MACE [19, 20], it has the advantage of accounting for the charge distribution. Presumably, this allows a high level of accuracy to be achieved, which results in amean absolute error (MAE) of just 30meVatom$^{-1}$ [17, 22]. Moreover, benchmarking against a large dataset, consisting of 257 thousand unrelaxed structures, has shown that CHGNet is one of the two best performing MLIPs currently available when compared based on multiple metrics [22].

All structures were relaxed, until all interatomic forces were below 0.001eVÅ$^{-1}$, using the BroydenFletcherGoldfarbShanno (BFGS) optimization algorithm, as implemented in the ASE toolkit [23], through functionalities provided in the CHGNet python package. In addition, the phonopy [24] code was utilized to determine harmonic force constants, which were, in turn, used to estimate the phononic contribution to the heat capacity. The training of the former, specifically, involved generating supercells with a minimum size of 12.5Å along each crystal axis and then systematically displacing selected atoms by 0.02Å and calculating the resulting forces. In all cases, the grid in reciprocal space was chosen to obtain a $q$-point density of 1000000atom$^{-1}$.

## B. Step 2: Training the model

### *1. Autoencoder structure*

A key component of the AL framework is the generative model that produces new candidate materials, here in the form of a WAE, which is able can encode known data into a structured latent space, while enforcing smooth latent manifolds [25]. In addition, it allows structural descriptors to be integrated with property predictions and, hence, allows the intricate dependence of the physical properties on the composition to be reconstructured, thereby ensuring that the design space is efficiently explored.

As both compositional and property information is fed to the model, it can learn joint representations, which enhances the predictive accuracy and interpretability. To be specific, the input not only includes a set of scalar descriptor, which represent the target properties, but also a pair of seven-dimensional composition vectors with normalized fractional occupation of the A- and B-site in the $ABO_3$ perovskite, respectively.

It is worth emphasizing that tests have revealed that multiple measures can be optimized simultaneously. In the present study, however, either the OC probability or OTC was considered, depending on the application.

As it is essential for the model to learn intricate nonlinear relationships, the encoder ($E_\theta$), which compresses the input vectors into an eight-dimensional latent space, has been structured as a series of layers, with decreasing size (80, 64, 48 neurons), that are all fully connected. While dropout regularization (10%) is only applied to the first of the hidden layers, to prevent overfitting; enhance robustness; and improve the generalization performance, all are accompanied by layer normalization (LN) and rectified linear unit (ReLU) activation functions to ensure high stability and convergence during training.

The use of a dual-decoder, which are designed to reconstruct separate parts of the input, is a distinctive feature of the current WAE architecture. Specifically, $LD_\theta$ has the task of take latent vectors and translate them into fractional occupations of the A- and B-sites. To maintain consistency with the encoder, both share the same layer structure (i.e., 48; 64; and 80 neurons). In addition, a softmax normalization is applied for each sublattice, which ensures valid and interpretable fractional compositions by enforcing physically meaningful compositional constraints. The property decoder ($PD_\theta$), on the other hand, has been scaled by a factor of four, resulting in larger fully connected layers (192, 256, 320 neurons). This is necessitated by the need to capture complex, nonlinear relationships between the latent space representation and the, scalar, physical properties, which places a higher demand on the representational capacity.

### *2. Wasserstein characteristic*

A crucial feature of the WAE is the loss function, which represents the training objective that is applied between the original and reconstructed vectors. In the present case, it includes a reconstruction term ($L_{recon}$) and a maximum mean discrepancy (MMD) penalty ($L_{MMD}$),

$$L_{total} = L_{recon} + \lambda \cdot L_{MMD}, \tag{1}$$

where $\lambda \in [0,1]$ controls the balance between the two. Specifically, $L_{recon}$ measures how accurately the decoder reconstructs compositions from latent embeddings while $L_{MMD}$ enforces alignment of the encoded latent distribution with the predefined prior. As the former is a $L_1$ loss, it provides balanced sensitivity to all vector dimensions and robustness against outliers, making it particularly well suited for compositional data. The other core component is the prior distribution, here represented by an isotropic Gaussian distribution ($N(0,I)$). Together with the aforementioned losses, it affords the "Wasserstein" characteristic and encourages the latent manifold to be smoothly varying as well as continuous, which is beneficial during the sampling and interpolation tasks in the AL cycle.

### *3. Gaussian mixture model*

The trained WAE is coupled with a GMM that is fit to the latent encoding. This allows a richer and more flexible latent space to be attained, which is capable of capturing complex compositional relationships and enabling controlled sampling of points likely to produce chemically reasonable compositions close to the training data manifold. By conditioning on known materials or regions associated with desirable properties, the generative model can be steered toward promising new candidates, which is especially useful in large compositional spaces.

## C. Step 3: Sample the latent space

### *1. Surrogate models*

A surrogate model is maintained that predicts the properties of candidate materials and quantifies the predictive uncertainty. Here, two different classes are employed, which are both capable of both property predictions and uncertainty estimations (UEs). The first involves adding a simple multi-layer perceptron (MLP), which accepts latent-space encodings as input and yields property predictions as output, to the WAE combined with MC dropout [26] to provide a heuristic UEs at the time of inference. Although this method is simple and scalable, it also has a tendency to become overconfident for out-of-distribution samples as it does not explicitly try to capture any underlying distributions within the data [27].

GP [28] represents an alternative which, in contrast to a MLP, predicts properties directly from the composition vectors, rather than encodings thereof. In addition, it is often better at identifying unexplored regions, and therefore guiding the AL cycle, thanks to more accurate uncertainty estimates. Here, the implementation in GPyTorch [29] was utilized, based on a Matérn-kernel [28] with individual length scales and an added scaling parameter. The hyperparameters were, more precisely, found by using an Adam optimizer [30] with negative marginal log likelihood as loss.

### 2. *Bayesian optimization*

AL combined with Bayesian optimization (BO) provides a systematic way to iteratively refine surrogate models of expensive-to-evaluate objective functions that has shown promise in guiding computational and experimental studies focused on materials discovery [9, 31, 32]. In essence, it is based on the core idea of balancing exploration and exploitation [33–35]. This concept is central for the training of the surrogate model, which begins with an initial introduction to a limited dataset, followed by successive updates after each round of querying new samples. The selection is more precisely guided by BO, using an acquisition function that picks points, from the predictive posterior of the model, that are either predicted to be optimal (exploitation) or likely to improve knowledge of the search space (exploration).

In each iteration of the AL, the acquisition sampling determines which candidates are to be evaluated, utilizing predictions and uncertainties to explore the design space efficiently [36]. For the present case, this process involves the generation of $n$ samples in each iteration, which corresponds to the number of parallel processes used for the evaluation. Each candidate is treated them as normally distributed random variables with a mean ($\mu$) and variance ($\sigma$) given by the prediction and uncertainty produced by the surrogate model. In particular, three different types of acquisition functions were tested: exploratory, Greedy, and Thompson sampling [37]. To achieve fair comparisons, independent of the surrogate model, candidates were randomly chosen from the composition space, which ensured that any improvements could be directly linked to their effectiveness in utilizing the predictions ($\mu$) and uncertainty estimates ($\sigma$).

### 3. *Explorative sampling*

Explorative sampling is a strategy that focuses on candidates with the largest uncertainty in the surrogate model predictions. The reason behind this is that by concentrating evaluations on regions where the model is least certain, one can improve the coverage of the design space and accelerate convergence to a reliable predictor. For the present case, this process involves selecting, at each iteration, the candidates whose uncertainty estimates ($\sigma$) are maximal, regardless of their predicted target value ($\mu$). Formally, this corresponds to choosing

$$\text{argmax}_x^{(n)}[\sigma], \tag{2}$$

where $n$ again denotes the number of parallel evaluations. This strategy emphasizes exploration of under-sampled regions and can be especially effective in the early stages of an active learning cycle, when model extrapolation is common.

### 4. *Greedy sampling*

Greedy sampling exclusively targets candidates with the highest predicted values of the property of interest, without explicitly accounting for uncertainty [38]. In this approach, the surrogate model predictions ($\mu$) are treated deterministically, and the top $n$ candidates are selected according to their expected performance,

$$\text{argmax}_{(x^{n})}[\mu]. \tag{3}$$

This strategy favors exploitation over exploration and can be highly effective once the model has reached sufficient maturity, allowing predictions to be reliable across the relevant design space. However, it risks premature convergence if applied too early, as it may overlook promising regions with high uncertainty.

### 5. Thompson sampling

Thompson sampling combines both exploration and exploitation, while remaining simple to understand and easy to implement [39, 40]. Instead of selecting candidates based only on their predicted mean values ($\mu$) or uncertainties ($\sigma$), this method samples from the full posterior distribution provided by the surrogate model. In each iteration, *n* candidate values are more precisely drawn from the normal distribution $N(\mu,\sigma)$ defined by the prediction and uncertainty of the surrogate model. From these samples, the candidate with the highest drawn value is selected:

$$\text{argmax}_x^{(n)}[f \sim \mathcal{N}(\mu, \sigma)]. \tag{4}$$

This stochastic process ensures that both high-performing regions (exploitation) and highly uncertain regions (exploration) are sampled according to their likelihood of providing an improvement.

## D. Step 4: Evaluate the properties

### 1. Vacancy formation energy

A special characteristic of perovskite oxides, compared to more conventional OCs, is the ability to release $O_2$ via the formation of vacancies. As such, the energy changed associated with a given transition $\delta_1 \rightarrow \delta_2$, such that $\delta_1 < \delta_2$, is essential for determining the oxygen transfer capacity of the material and, thus, the usefulness in CL applications. Specifically, the corresponding reaction,

$$\text{Sr}_{1-x}\text{A}_x\text{Fe}_{1-y}\text{B}_y\text{O}_{3-\delta_1} \rightleftharpoons \text{Sr}_{1-x}\text{A}_x\text{Fe}_{1-y}\text{B}_y\text{O}_{3-\delta_2} + \frac{\delta_2\text{-}\delta_1}{2}\text{O}_2, \tag{5}$$

is associated with a free energy,

$$\Delta G_{\delta_1 \rightarrow \delta_2} = G_f^{\delta_2} + \frac{\Delta\delta}{2}\mu_{\text{O}_2} - G_f^{\delta_1}, \tag{6}$$

where $G_f^{\delta_i} = G_f\,(\text{CaMnO}_{3-\delta_i})$ and $\Delta\delta = \delta_2 - \delta_1$.

By assuming that the volume contribution, $PV_{\delta_i}^{0\,K}$, is negligible, the enthalpy at 0K can be approximated by the internal energy $H^{\delta_i}(0\text{K}) \approx U^{\delta_i}(0\text{K})$. Furthermore, the difference between the specific heat capacity at constant volume and pressure was neglected, thus allowing the formation energies, at any temperature *T*, to be calculated as

$$G_f^{\delta_i}(T) = H_f^{\delta_i}(T) - TS_f^{\delta_i}(T), \tag{7}$$

with the enthalpy and entropy are given by

$$H_f^{\delta_i}(T) \approx H_f^{\delta_i}(0\,\text{K}) + \int_0^T C_V^{\delta_i}(T')dT', \tag{8}$$

$$S_f^{\delta_i}(T) \approx S_f^{\delta_i}(0\,\text{K}) + \int_0^T \frac{C_V^{\delta_i}(T')}{T'}dT'. \tag{9}$$

### 2. Oxygen carrier probability

The capability of a specific material to act as an OC for either CLAS or CLDR was assessed by comparing the vacancy formation energies ($\Delta G_{\delta_1 \rightarrow \delta_2}$) obtained from the first principles calculations with the ranges defined by Wang et al. [11] (see table Table I). It is worth emphasizing that these intervals were chosen based on empirical knowledge regarding the pressure swings ($P_{O_2}^{min} \rightarrow P_{O_2}^{max}$) commonly employed in the respective processes. The material must, therefore, release oxygen within this range. In the case of perovskites, the vacancy formation energy should, therefore, satisfy the criteria

$$\Delta G_{\min} \leq \Delta G_{\delta_1 \rightarrow \delta_2} \leq \Delta G_{\max}, \tag{10}$$

$$\Delta G_{\min/\max} = -\frac{RT}{2} \ln \frac{P_{O_2}^{\min/\max}}{P^0} \tag{11}$$

where $T$ is the temperature, $R$ the gas constant, and $P^0$ the atmospheric pressure. One reason for selecting the aforementioned CL processes is that they represent opposite extremes since the viable range for the oxygen partial pressure range is much more narrow and closer to the normal value for CLAS ($\sim$ 0.01atm to 0.2atm) compared to CLDR ($\sim 10^{-21}$ atm to $10^{-17}$ atm). The reason for this disparity is that the goal of the former is to generate $O_2$ while the latter is used to produce syngas (CO and $H_2$), which, naturally, has a significant impact on the conditions within the reduction reactors. In addition, the OC is regenerated with $CO_2$ in the case of CLDR while air is used in CLAS, as well as CLOU.

TABLE I. Ranges in $\Delta G$ that are deemed ideal for CLAS and CLDR OCs together with the extended limits used, by by Wang et al. [11], in their screening procedure. In the present study, the latter were employed to estimate the probability that a given candidate is suitable for the respective CL processes.

| Process | Range | Temperature | $\Delta G_{min}$ | $\Delta G_{max}$ |
|---|---|---|---|---|
| CLAS | $\sim$ 0.01atm −−0.2atm | 400°C | 0.05eV | 0.13eV |
| CLAS | $\sim$ 0.$_{21}$01atm −−0.2$_{17}$atm | 700°C | 0.07eV | 0.19eV |
| CLDR | | 800°C | 2.23eV | 2.24eV |
| CLDR | $\sim$ $10^{-21}$ atm −−$10^{-17}$ atm $\sim$ $10^{-}$ atm −−$10^{-}$ atm | 950°C | 2.13eV | 2.41eV |
| CLAS | Extended* | 400°C | −0.25eV | 0.63eV |
| CLAS | Extended* | 700°C | −0.23eV | 0.69eV |
| CLDR | Extended* | 800°C | 1.93eV | 2.74eV |
| CLDR | Extended* | 950°C | 1.83eV | 2.91eV |

* An extended $\Delta G$ range was used by Wang et al. [11] for screening OCs to compensate for the discrepancy between experimental and calculated vacancy formation energies.

Since Wang et al. [11] required strict limits for their high-throughput screening process, they found it necessary to extend the aforementioned intervals due to the significant deviation between computed and experimental formation energies. The goal of the study at hand, however, was to identify the best material for each specific application, rather than as many potential candidates as possible. Consequently, it was necessary to construct an appropriate measure for guiding the AL process. A natural choice is the probability that a subset of the vacancy formation energies, for a given composition, is found within the aforementioned limits, at one or both temperatures. To allow such a comparison, each energy was treated as a randomly distributed variable $X \sim N(\mu, \sigma^2)$ with a mean ($\mu$) given by the value obtained from the first principles calculations and a standard deviation of $\sigma$ = 0.4eV. It should be noted that the latter value was selected based on a comparison of the widths of the original and extended intervals proposed by Wang et al. [11]; the mean was also shifted by the difference between the respective centres.

While calculations were only performed for structures with $\delta \in \{0, 0.125, 0.25, 0.375, 0.5\}$, the formation energies at intermediate values, i.e. $\delta \in \{0.0625, 0.1875, 0.3125, 0.4375\}$, were obtained via interpolation, as in the study by Wang et al. [11]. Furthermore, the materials were compared based on the probability $P(\Delta G_{min} < X \leq \Delta G_{max})$ that the energy associated with any of the possible transitions $\delta_1 \rightarrow \delta_2$, such that $\delta_2 - \delta_1 \equiv \Delta\delta = 0.125$, were found within the ideal limits ($\Delta G_{min}$, $\Delta G_{max}$) defined by Wang et al. [11] (see table Table I). In particular,

$$P\left(\Delta G_{min} < X \leq \Delta G_{max}\right) = F_X(\Delta G_{min}) - F_X(\Delta G_{max}), \tag{12}$$

where

$$F_X(x) = F(x;\mu,\sigma) \\ = \frac{1}{\sigma\sqrt{2\pi}} \int_{-\infty}^{x} \exp\left(-\frac{(t-\mu)^2}{2\sigma^2}\right) \mathrm{d}t \tag{13}$$

is the cumulative distribution function for a normally distributed variable.

### 3. *Oxygen transfer capacity*

The OTC was deemed to be the most appropriate measure of the CLOU capability, mainly because it is widely used within the chemical looping combustion (CLC) community and is relatively straightforward to determine experimentally via, i.e., thermogravimetric analysis. This involved comparing the formation energies for the different vacancy concentrations, obtained from the first principles calculations, at oxidizing ($p_{O_2} = 0.2\text{atm}$, $T = 950$ °C) and reducing ($p_{O_2} = 10^{-14}$ atm, $T = 1050$ °C) conditions, which represent the environments within the air and fuel reactors, respectively. Subsequently, the OTC was calculated as $\text{OTC} = 0.5\Delta\delta M_{O_2}/M_{ABO_3}$, where $\Delta\delta = \delta_{red} - \delta_{ox}$ represents the non-stoichiometry difference and $M_i$ the molar mass of compound $i$.

## III. RESULTS AND DISCUSSION

This section begins with a brief comparison of the training data used in this study and by Wang et al. [11], respectively. Thereafter, the main results of the calculations outlined in section II are presented and discussed, beginning with those pertaining to OCs for CLDR, followed by CLAS and, finally, high entropy oxygen carriers (HEOCs).

### A. Validation of training data

Although the vacancy formation energies obtained with CHGNet [17] differ from the those reported by Wang et al. (see Figure 2 and figure 6 in [11]), the overall agreement is fair. The tendency for CHGNet to give lower values than those reported by Wang et al. [11] can be attributed to a number of factors. Specifically, the former potential has been trained on DFT energies which have not only been adjusted, to provide a better agreement with experimental data, but also performed with other settings. Evidently, Wang et al. [11] did not include such a correction since it had to be artificially removed when using CHGNet in order to obtain similar predictions. To check that this procedure did not have any spurious effects, the calculations outlined in section IIA3 were repeated with the M3GNet [18] (see Figure 3), which is based on pristine DFT energies from the Materials Project database. As expected, it was found that the uncorrected vacancy formation energies obtained with the respective MLIPs agreed almost perfectly.

### B. Chemical looping dry reforming

As is explained by Wang et al. [11], the CLDR process places rather strict requirements on the OCs, making the task of finding suitable materials difficult. It would, therefore, be expected that this is a problem for which the previously outlined AL strategy would be well suited. This hypothesis is supported by a comparison of the counts of the number of candidates identified using a variety of different approaches (Figure 4).

The initial 10% subset selected from the full dataset, which was used to train the surrogate model, has a similar pattern as the overall distribution of probabilities (notice the log scale in Figure 4a). To be specific, most candidates are concentrated at values close to zero, while only a few are in the higher-probability range. Similar results were obtained when applying random sampling—most candidates cluster at very low probabilities, and only a small number appear in the higher range—confirming that unguided selection is an unwise strategy for materials with a vast composition space (see Figure 4b). Explorative sampling, by contrast, shifts the distribution somewhat, though not in a way that favors the discovery of promising OCs. By construction, this strategy focuses on regions of the latent space where the model is most uncertain. This does expand coverage but also leads to fewer high-probability candidates, although slight less so when using MC dropout method for UE compared to GP.

A greedy strategy could in principle be more effective, in particular if the entire phase space has been mapped or all the best materials are located within the surveyed area. It is therefore quite interesting that much fewer data points are added when such an approach is combined with GP compared to MC dropout (see Figure 4e,f). Since this might initially seem to favor the latter

UE method, the opposite is in fact true. The reason is that a greedy strategy, if efficiently implemented, is bound to get trapped, as it is designed to only focus on the region that, based on the available information, is the most likely to contain the best candidates. Hence, combining it with MC dropout apparently leads to a more uncertain model that consequently samples a larger area and, thus, adds more points. The clearest improvement arises from Thompson sampling (see Figure 4g,h). By drawing from the full posterior distribution, this strategy achieves a balance between exploration and exploitation. As a result, the distributions are strongly skewed toward a high probability and generate far fewer candidates on the lower end of the scale than random sampling, pure exploration or a greedy strategy. The resulting data are instead centered around a higher probability, close to $P \approx 0.04$. In practice, the combination of Thompson sampling and GP emerges as the most efficient approach for discovering promising OCs for CLDR. Even so, it is worth noting that MC dropout is nearly as effective for generating UEs, as the upper end of the distribution reaches comparably high values.

While the aggregated data revealed distinct differences in efficacy depending on the strategy, these trends become even clearer when considering the evolution of the CLDR probability (see Figure 5). For instance, a greedy strategy combined with GP initially identifies many high-performing materials relatively quickly, but ceases to find any new compositions—only duplicates—after just 12 cycles (see Figure 5b,e). When the uncertainty is estimated via MC dropout, meanwhile, multiple data points with a relatively high average ($P \approx 0.030$) are added in each cycle. A very similar trend is observed when using Thompson sampling, but in this case the performance actually improves markedly when switching to GP (see Figure 5c,f). Specifically, doing so means that the average probability not only rises more sharply in the beginning but also flattens out at a higher value ($P \approx 0.035$). It is also worth mentioning that while exploration generates candidates with a much lower mean, the maximum value in each cycle is close to those obtained with the other approaches (see Figure 5a,d). Consequently, there exist promising candidates within unexplored areas of the latent space, which explains why Thompson sampling, which balances exploration and exploitation, achieves better results than a greedy strategy.

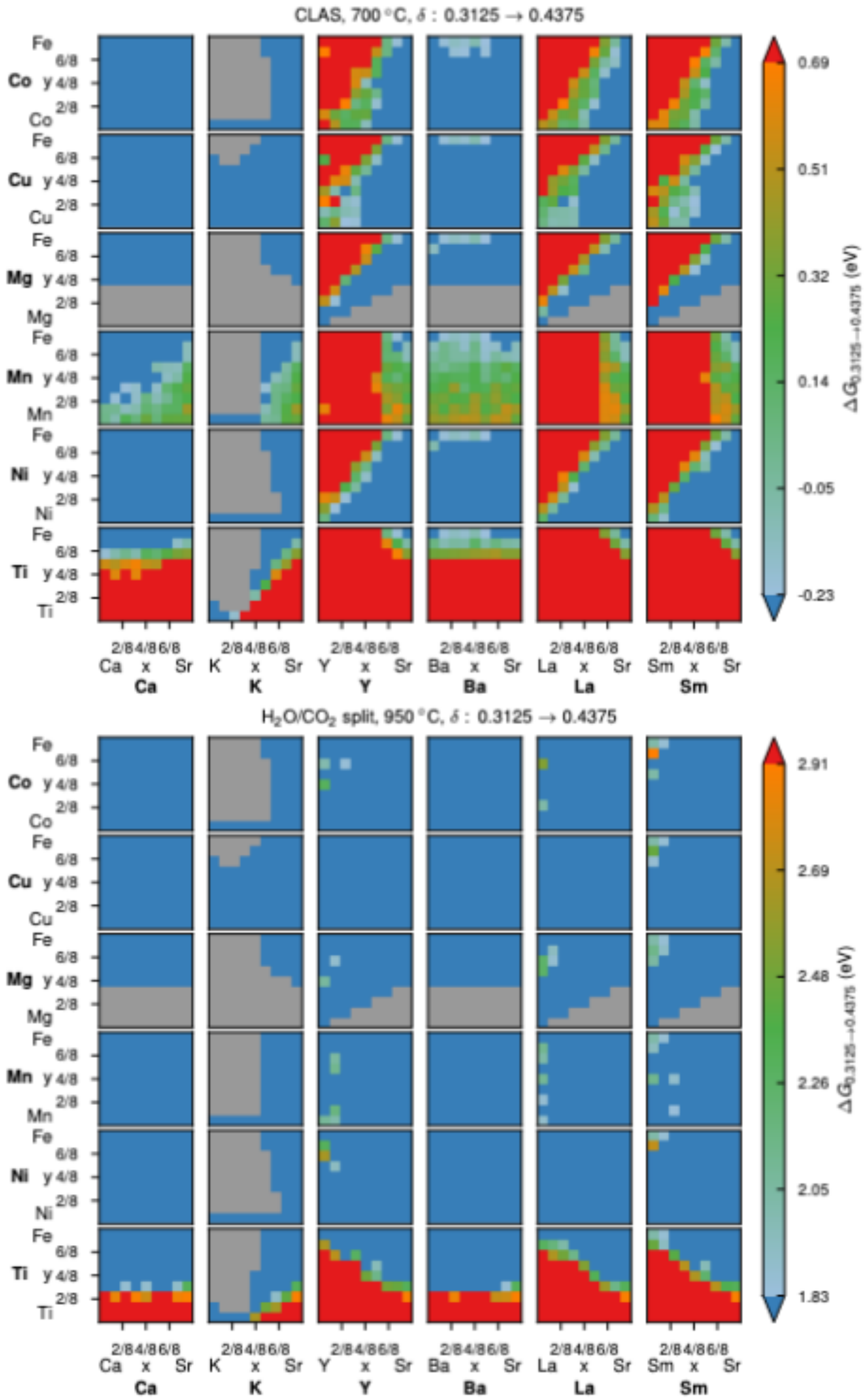

FIG. 2. Heatmaps with $\Delta G_{0.3125\to0.4375}$ calculated using [17] for all $Sr_{1-x}A_xFe_{1-y}B_yO_{3-\delta}$ configurations, where A = Ca, K, Y, Ba, La, Sm; B = Co, Cu, Mn, Mg, Ni, Ti; and $x, y \in \{0, 0.125, 0.25, 0.375, 0.5, 0.625, 0.75, 0.875, 1\}$. Specifically, values within the appropriate ranges for (a) CLAS at 700°C ($-0.23\text{eV} \le \Delta G_{0.3125\to0.4375} \le 0.69\text{eV}$) and (b) CLDR at 950°C ($1.83\text{eV} \le \Delta G_{0.3125\to0.4375} \le 2.91\text{eV}$) have a turquoise-green-orange shade while those above and below are plain blue and red, respectively. Note that these diagrams are comparable with figures 6a and 6c in Wang et al (2022) [11].

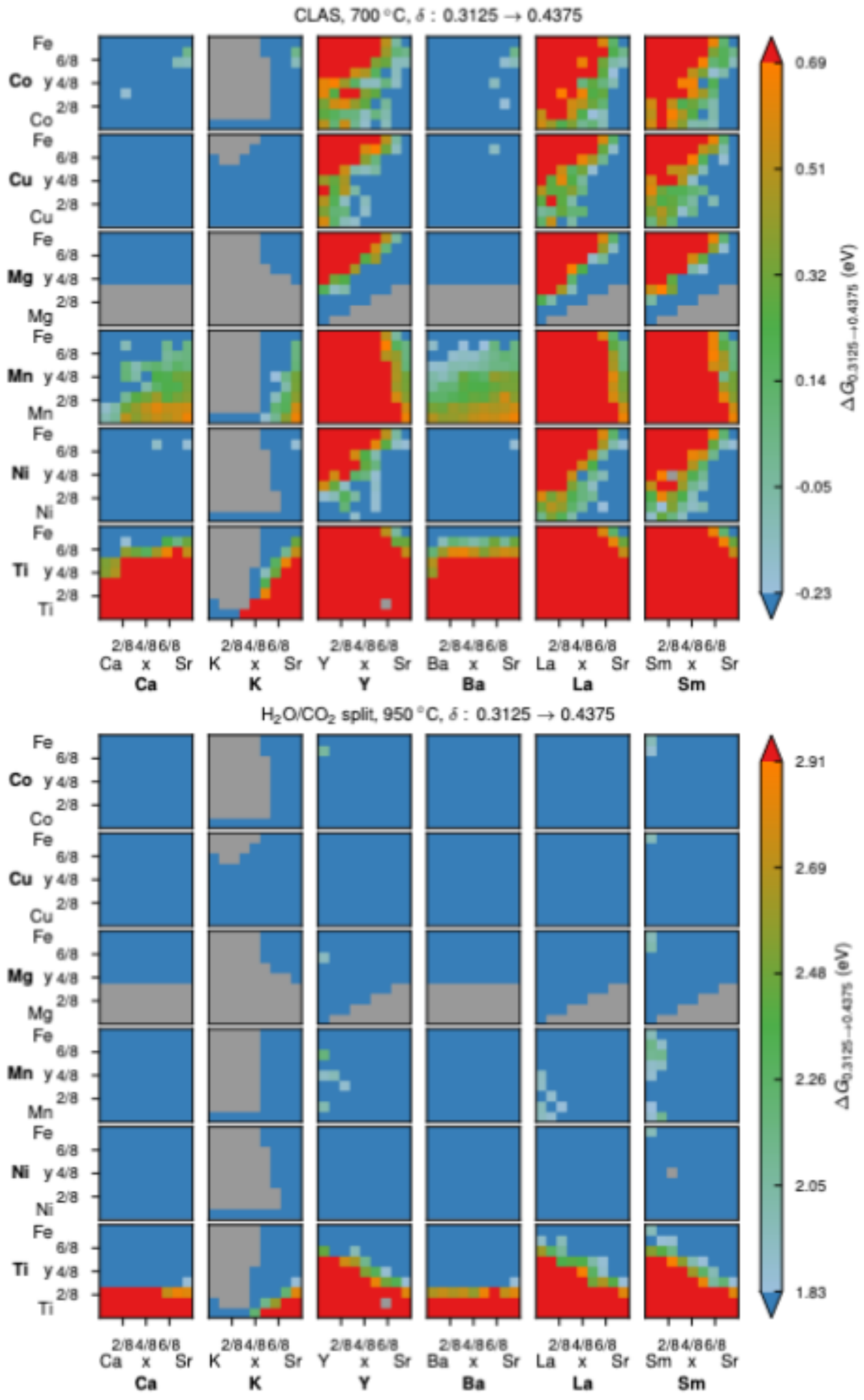

FIG. 3. Heatmaps of $\Delta G_{0.3125\rightarrow0.4375}$ calculated using [18] for all $Sr_{1-x}A_xFe_{1-y}B_yO_{3-\delta}$ configurations, where A = Ca, K, Y, Ba, La, Sm; B = Co, Cu, Mn, Mg, Ni, Ti; and $x, y \in \{0, 0.125, 0.25, 0.375, 0.5, 0.625, 0.75, 0.875, 1\}$. Specifically, values within the appropriate ranges for (a) CLAS at 700°C ($-0.23\text{eV} \leq \Delta G_{0.3125\rightarrow0.4375} \leq 0.69\text{eV}$) and (b) CLDR at 950°C ($1.83\text{eV} \leq \Delta G_{0.3125\rightarrow0.4375} \leq 2.91\text{eV}$) have a turquoise-green-orange shade while those above and below are plain blue and red, respectively. Note that these diagrams are comparable with figures 6a and 6c in [11].

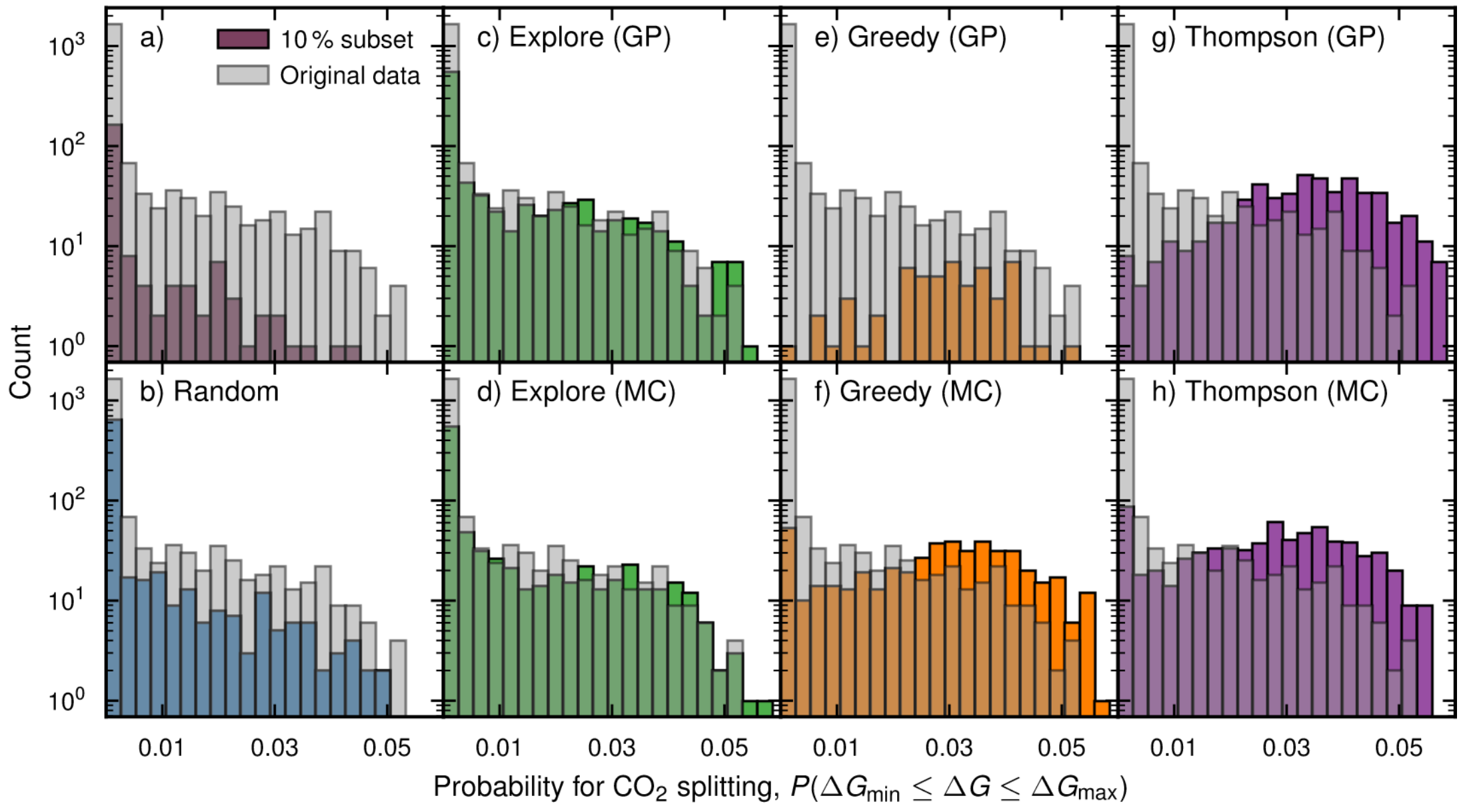

FIG. 4. Histograms based on the probability that a given OC candidate is useful for CLDR showing the original dataset (light grey) together with the 10% subset used for training (a) and the results obtained with different AL approaches. Specifically, this includes random selection (b) as well as sampling based on exploration (c,d), a greedy strategy (e,f), and the Thompson method (g,h) combined with either GP (c,e,g) or MC dropout (d,f,h). Note that all panels, except (a), show results obtained from five individual runs after duplicate entries have been removed.

An interesting observation is that the configurational entropy increases continuously for most method combinations, even though this property is not included in the training of the WAE (see Figure 5g-l). A likely reason, however, is that the original training data is not optimally constructed for AL. More precisely, it was constructed in compliance with the study by Wang et al. [11] and, thus, only covers a small portion of the latent space, namely binary; tertiary; and quaternary oxides obtained via by substitution of Fe and Sr in $Sr_{1-x}A_xFe_{1-y}B_yO_{3-\delta}$. As such, the increasing trend speaks to the inherent robustness of AL, especially when combined with an effective sampling method, and indicates an inherent tendency to discover HEMs, in spite of the fact that no specific restrictions or conditions have been applied to ensure such an outcome. The average abundance of the metallic element, meanwhile, remains relatively constant, even though a decrease would be expected since the added candidates should successively be made up of a higher number of distinctive, and therefore rarer, atomic species. In fact, this is only observed when using Thompson sampling in combination with GP, which therefore seems to explore other regions of the latent space.

As the models are trained to maximize the CLDR probability, the average compositions should give an indication of which elements are the most likely to provide favorable qualities (see Figure 6). In this context, it is worth remembering that Sr and Fe are the most common species in the original data set, since the latter was constructed through elemental substitutions in $Sr_{1-x}A_xFe_{1-y}B_yO_{3-\delta}$ (see Figure 6i-j). For a random sampling, however, all elements appear with about the same frequency except Mg and, to a lesser extent, K (see Figure 6g-h). As mentioned by Wang et al. [11], the reason is that the former (latter) is the only alkali (alkaline earth) metal on the A (B) site, which means that it is less likely to form stable perovskite structures, based on the Bartel tolerance factor as well as the requirement of charge neutrality (see Figure 2). The AL models, on the other hand, seem to favor La followed by Sr, Sm, and Y together with Mn as well as Fe and, to some extent, Co (see Figure 6a-f). Even so, there are quite prominent differences when comparing the individual approaches. As anticipated, exploration leads to a more uniform distribution, even if the trends mentioned earlier are still visible. Another observation is that greedy sampling based on GP gives a narrow spread of candidates that primarily contain La and Fe, which is indicative of the inherent "nearsightedness". Yet when combined with MC dropout, the average concentrations of the former two elements drop slightly to the benefit of Y, Sr, and Sm, in the former case, as well as Mn and Co, in the latter. Surprisingly, the values remain more or less the same for the Thompson method. Slight shifts can be observed, however, if a GP-based UE is used, which yields more La, Mn, and Co but less Y, Sm, Sr, and Fe.

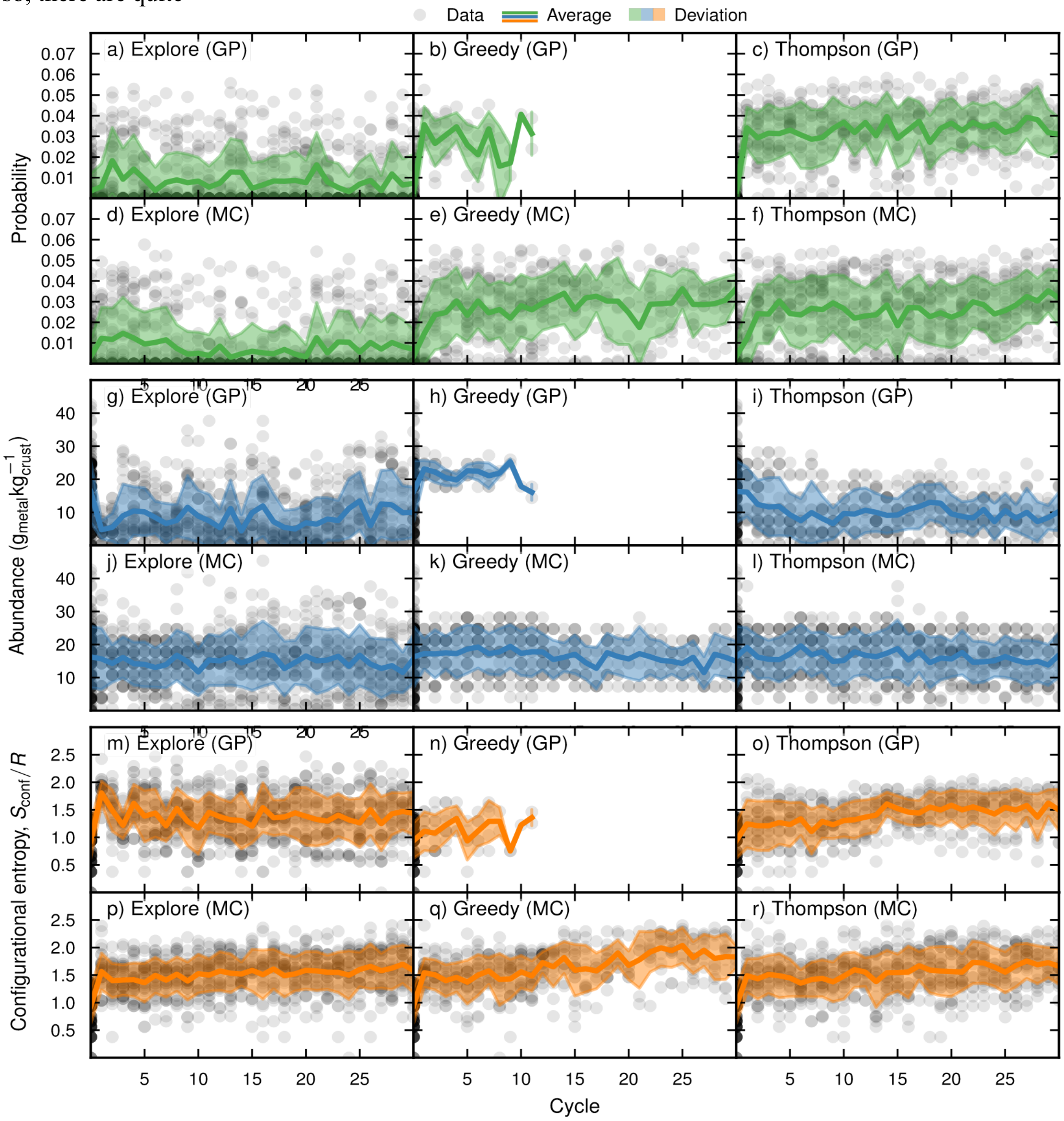

FIG. 5. Scatter plots with the probability that a given OC candidate is useful for CLDR (a-f); average abundance of the metallic elements in the earth's crust (g-l); and configurational entropy, $S_{conf}/R$, (m-r) in the original dataset (light grey) as well as those added in each cycle (purple) during five individual runs, after removing duplicates. The corresponding averages (solid line) and standard deviations (filled curve) are also shown.

In conclusion, it would seem that an ideal candidate is likely to be based on $LaMnO_3$, where La (Mn) has, in part, been substituted by Sr and Sm (Fe and Co). It could be argued that the relatively high percentage of Fe and Sr stems from the skewness of the original training data. This appears less likely, however, given the formulations that appear at the top of the leaderboard; obtained via a combination of Thompson sampling and GP (see Table II). In fact, Fe;

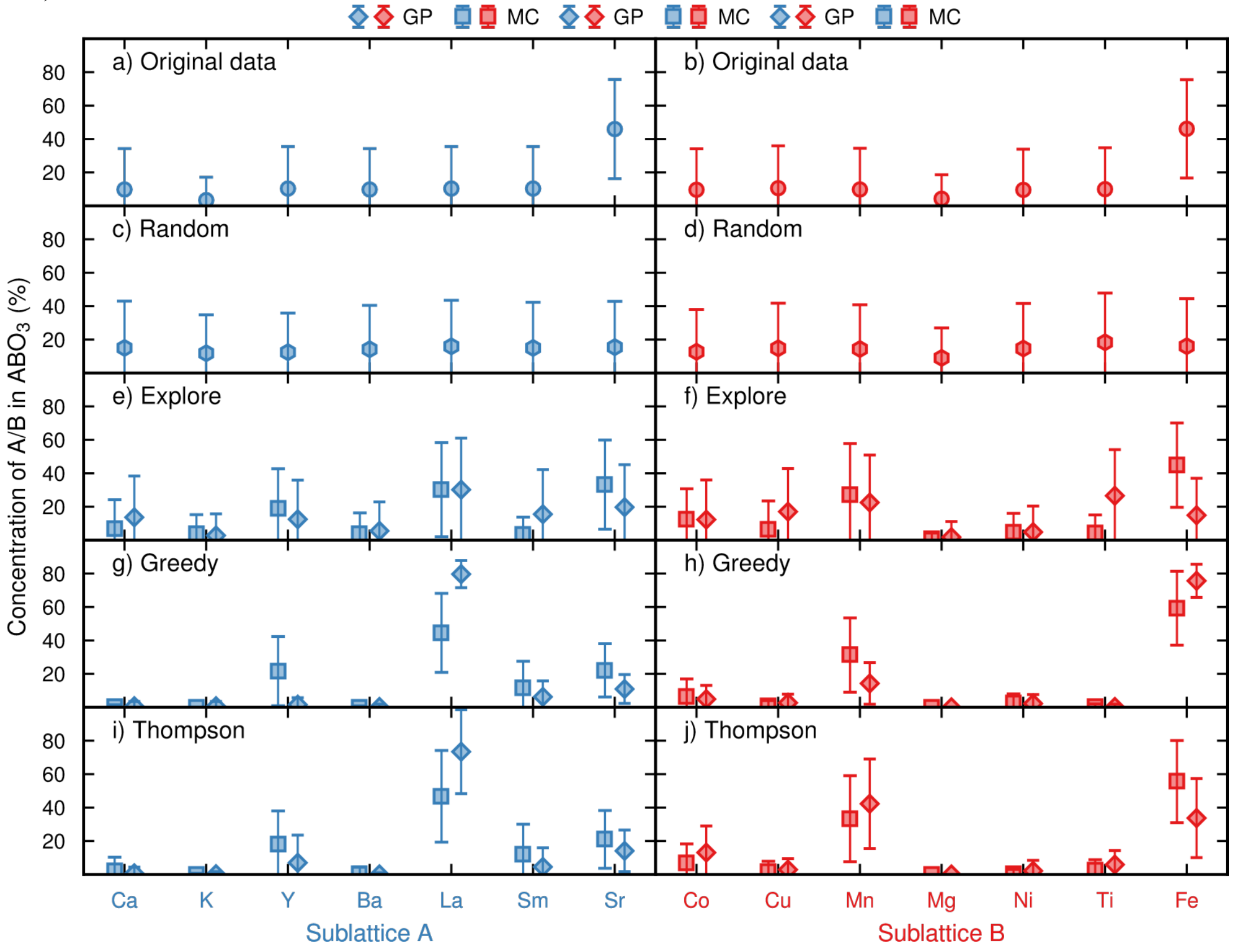

FIG. 6. Diagrams with average counts of elements located on sublattice A (a,c,e,g,i) and B (b,d,f,h,j) in $ABO_3$ OCs, where the bars indicate the corresponding standard deviations, for CLDR. Specifically, this includes results from sampling based on exploration (a-b); a greedy strategy (c-d); and the Thompson method (e-f), combined with either GP (diamonds) or MC dropout (squares), as well as random selection (g-h) and the original dataset (i-j).

Sr; Ti; and Co are, respectively, found in as many as 17; 16; 17; and 12 of the 20 best candidates, including the first four: $CoFe_4La_7Mn_3SrO_{24}$; $Co_3Fe_4La_7SrTiO_{24}$; $Fe_3La_6Mn_3NiSr_2TiO_{24}$; and $CoCuFe_2La_7Mn_3SrTiO_{24}$.

### C. Chemical looping air separation

While identical computational workflows and training datasets were applied in the search for suitable candidates for both CLAS and CLDR, the underlying optimal conditions vary significantly between the two processes. For the sake of brevity, however, the discussion that follows will focus on the areas where distinctive differences are observed. Beginning with the candidate counts after the final AL cycle, the trends are very similar to those described at the beginning of the previous section (see Figure 7). Nevertheless, there are two key differences. Firstly, the choice of UE seems to be inconsequential and, secondly, greedy sampling apparently outperforms the Thompson method, since the upper end of the distribution is slightly shifted towards higher probabilities. Still, neither is able to identify any new compositions with a significantly better performance than those in initial training set.

When considering the evolution over the cycles, the strong similarities in the efficacy of the alternative approaches are equally apparent (see Figure 8). In particular, the average CLAS probability remains almost constant throughout, albeit at a higher level when using a greedy strategy or Thompson sampling compared to exploration. If anything, the trend is somewhat decreasing. The metal abundance, on the other hand, displays a steady decline while the configurational entropy increases, although more tentatively.

Given the aforementioned results, it is not astounding that the almost negligible variations are observed when comparing the compositions of the candidates (see Figure 9). With the exception of a somewhat broader distribution when using pure exploration, it is evident that Sr (Fe) followed by Ba (Co and Cu) are the most common elements on the A (B) site. Naturally, this holds true for the highest ranking candidates, among those sampled using the

| Place | Formula | Cycle | Probability | $S_{conf}/R$ | Metal abundance |
|---|---|---|---|---|---|
| 1st | $CoFe_4La_7Mn_3SrO_{24}$ | 17 | 0.0585 | 1.35 | 1.43e+04 |
| 2nd | $Co_3Fe_4La_7SrTiO_{24}$ | 14 | 0.0583 | 1.35 | 1.45e+04 |
| 3rd | $Fe_3La_6Mn_3NiSr_2TiO_{24}$ | 21 | 0.0583 | 1.82 | 1.12e+04 |
| 4th | $CoCuFe_2La_7Mn_3SrTiO_{24}$ | 3 | 0.0574 | 1.87 | 7.61e+03 |
| 5th | $Fe_4La_7Mn_3NiSrO_{24}$ | 24 | 0.057 | 1.35 | 1.43e+04 |
| 6th | $CoFeLa_5Mn_4Sr_3Ti_2O_{24}$ | 12 | 0.0564 | 1.87 | 4.55e+03 |
| 7th | $Co_4FeLa_8Mn_2TiO_{24}$ | 9 | 0.0561 | 1.21 | 4.02e+03 |
| 8th | $Co_5La_8Mn_2TiO_{24}$ | 20 | 0.0554 | 0.9 | 499 |
| 9th | $Co_2Fe_5La_7SrTiO_{24}$ | 17 | 0.0553 | 1.28 | 1.8e+04 |
| 10th | $Fe_3La_7Mn_2Ni_2SrTiO_{24}$ | 15 | 0.0553 | 1.7 | 1.11e+04 |
| 11th | $CoFe_2La_6Mn_4Sr_2TiO_{24}$ | 2 | 0.055 | 1.78 | 7.69e+03 |
| 12th | $CoFe_3La_7Mn_2NiSrTiO_{24}$ | 28 | 0.055 | 1.87 | 1.11e+04 |
| 13th | $Fe_4La_7MnNi_2SrTiO_{24}$ | 14 | 0.0547 | 1.59 | 1.45e+04 |
| 14th | $CaCo_3Fe_4La_7TiO_{24}$ | 13 | 0.0541 | 1.35 | 1.7e+04 |
| 15th | $FeLa_5Mn_5Sr_3Ti_2O_{24}$ | 14 | 0.054 | 1.56 | 4.6e+03 |
| 16th | $Co_4La_7Mn_3SrTiO_{24}$ | 11 | 0.0539 | 1.35 | 578 |
| 17th | $CaFe_2La_6Mn_6SrO_{24}$ | 12 | 0.0535 | 1.3 | 1e+04 |
| 18th | $Co_4La_8Mn_3TiO_{24}$ | 21 | 0.0533 | 0.974 | 557 |
| 19th | $FeLa_7Mn_4Ni_2SrTiO_{24}$ | 28 | 0.0533 | 1.59 | 4.16e+03 |
| 20th | $CuFe_3La_7Mn_3SrTiO_{24}$ | 1 | 0.0532 | 1.63 | 1.11e+04 |

TABLE II. Leaderboard with the top 20 OCs, found via Thompson sampling based on GP, that have been ranked by the probability that they would be suitable for CLDR. The table also includes data on the, first, cycle where the candidate was identified, the configurational entropy ($S_{conf}/R$), and the average metal abundance in the earth's crust ($g_{metal}\,kg^{-1}_{crust}$).

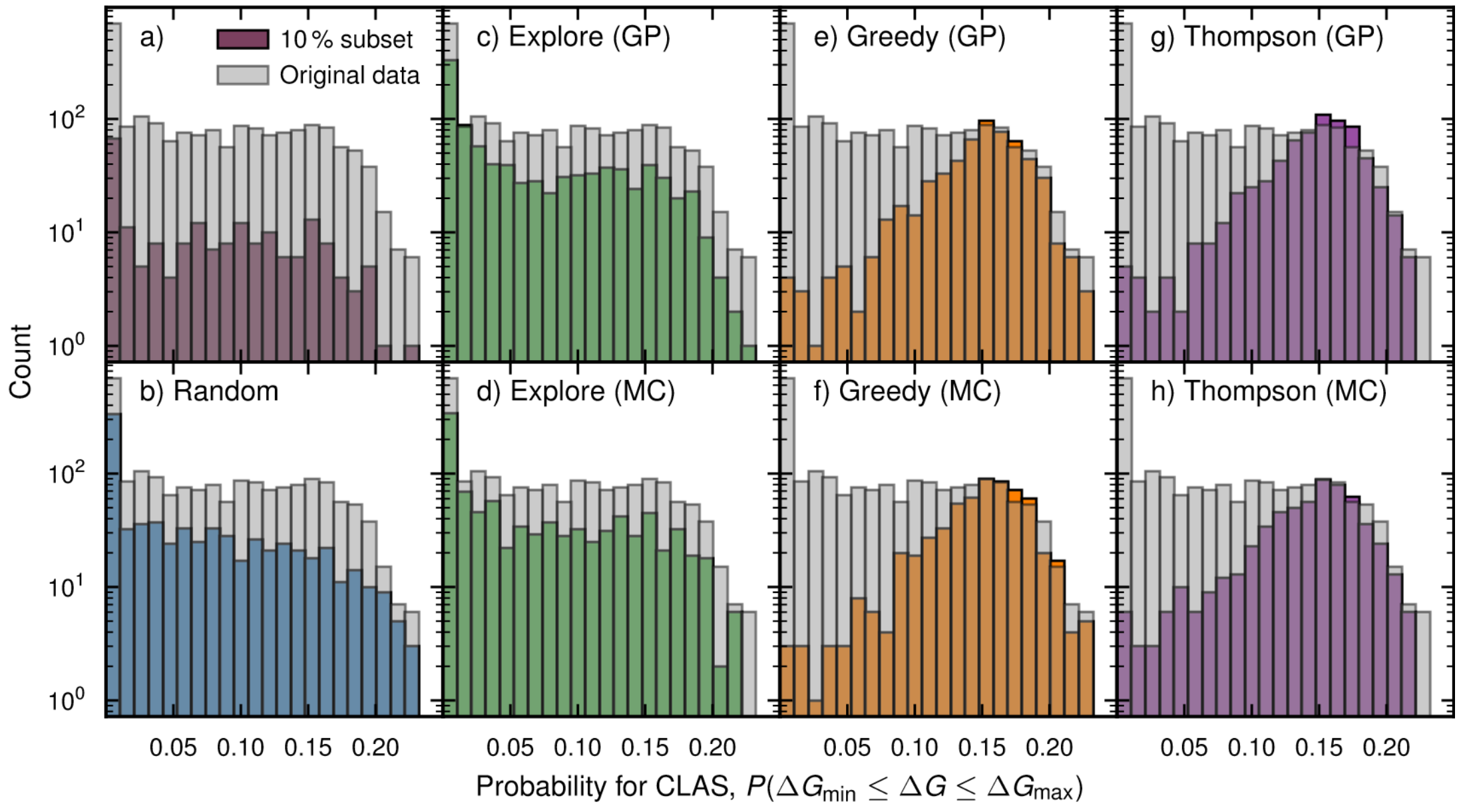

FIG. 7. Histograms based on the probability that a given OC candidate is useful for CLAS, showing the original dataset (light grey) together with the 10% subset used for training (a) and the results obtained with different AL approaches. Specifically, this includes random selection (e) as well as sampling based on exploration (b,f), a greedy strategy (c,g), and the Thompson method (d,h) combined with either GP (b-d) or MC dropout (f-h). Note that all panels, except (a), show results obtained from five individual runs after duplicate entries have been removed.

Thompson method combined with GP: $Co_6Fe_2La_2Sr_6O_{24}$, $Fe_6La_3Mg_2Sr_5O_{24}$, $Fe_5Mg_3Sm_4Sr_4O_{24}$, and $Ba_4Fe_8Sr_4O_{24}$ (see Table III). Interestingly, these four, as well as nine others among the 16 that follow, all stem from the initial set

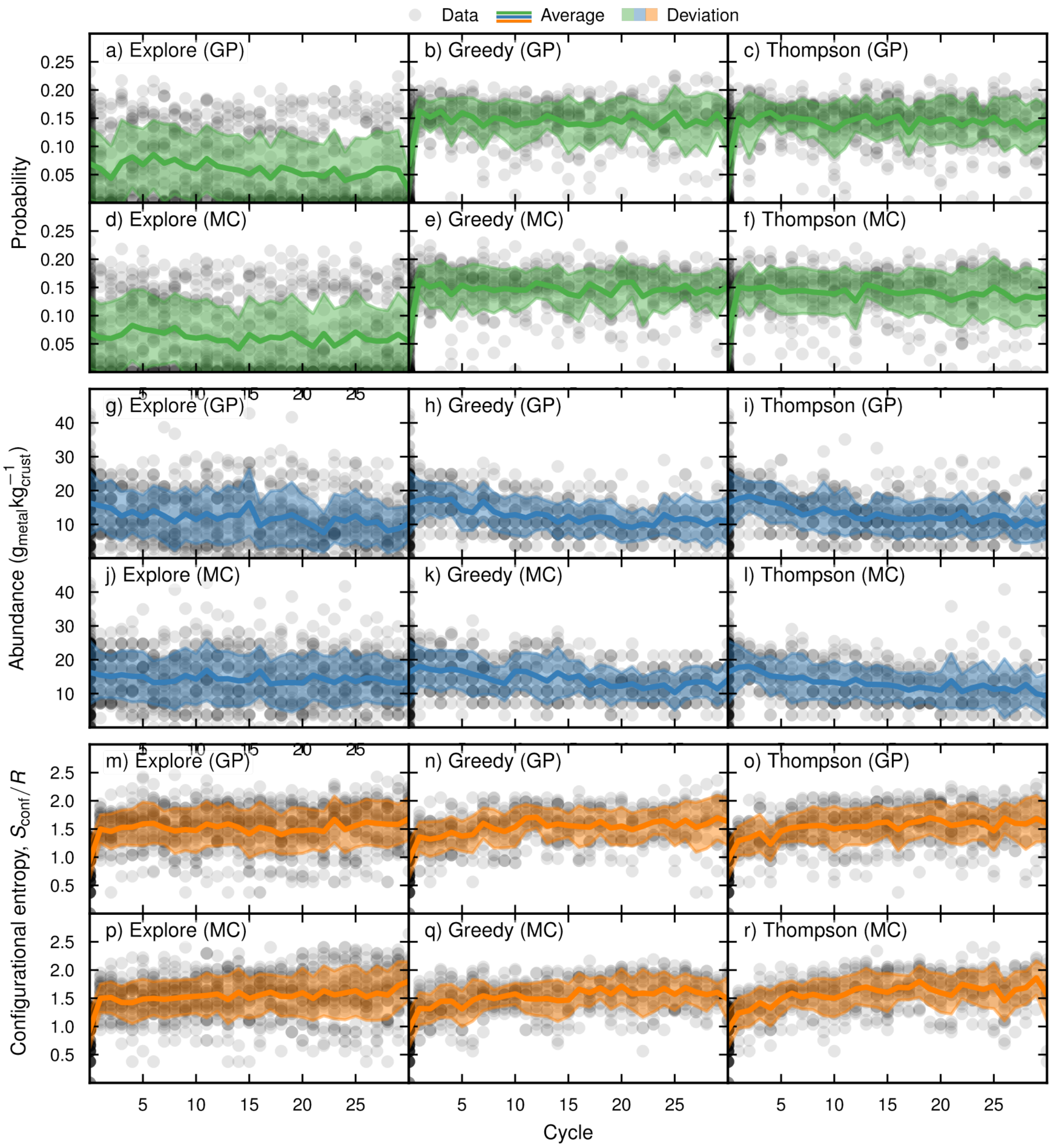

FIG. 8. Scatter plots with the probability that a given OC candidate is useful for CLAS (a-f); average abundance of the metallic elements in the earth's crust (g-l); and configurational entropy, $S_{conf}/R$, (m-r) in the original dataset (light grey) as well as those added in each cycle (purple) during five individual runs, after removing duplicates. The corresponding averages (solid line) and standard deviations (filled curve) are also shown. of $Sr_{1-x}A_xFe_{1-y}B_yO_{3-\delta}$ training structures. While this may, at first glance, appear as a failure, it should be noted that these candidates are correctly identified via AL when only a subset of the original data is used for training. The lack of improvement is therefore an indication that the problem of finding OCs based on the CLAS probability, at least when defined by such a broad energy interval, is not specific enough to warrant the utilization of advanced search methods.

While the AL seems to be more effective at discovering OCs CLDR than CLAS the opposite is true for the ML model trained by Wang et al. [11], which was able produce about four times as many viable candidates in the latter case. Taken together with the superiority of greedy sampling, this suggests that the problem of finding

CLAS OCs, at least based on the broad criteria utilized here and by Wang et al. [11], is handled equally well by conventional ML approaches. One should, nonetheless, remember that there exist other potential causes, such as a poor choice of initial training data and lax ranges for the vacancy formation energies. As advertised earlier, some blame might also fall on the choice of target property. To be specific, any probability-based measure will, by definition, have a maximum value, which, in the present case, is attained by members in the original training set. As such, reaching even higher values could be an impossible task and the apparent failure to do so should thus not be attributed to the AL strategies.

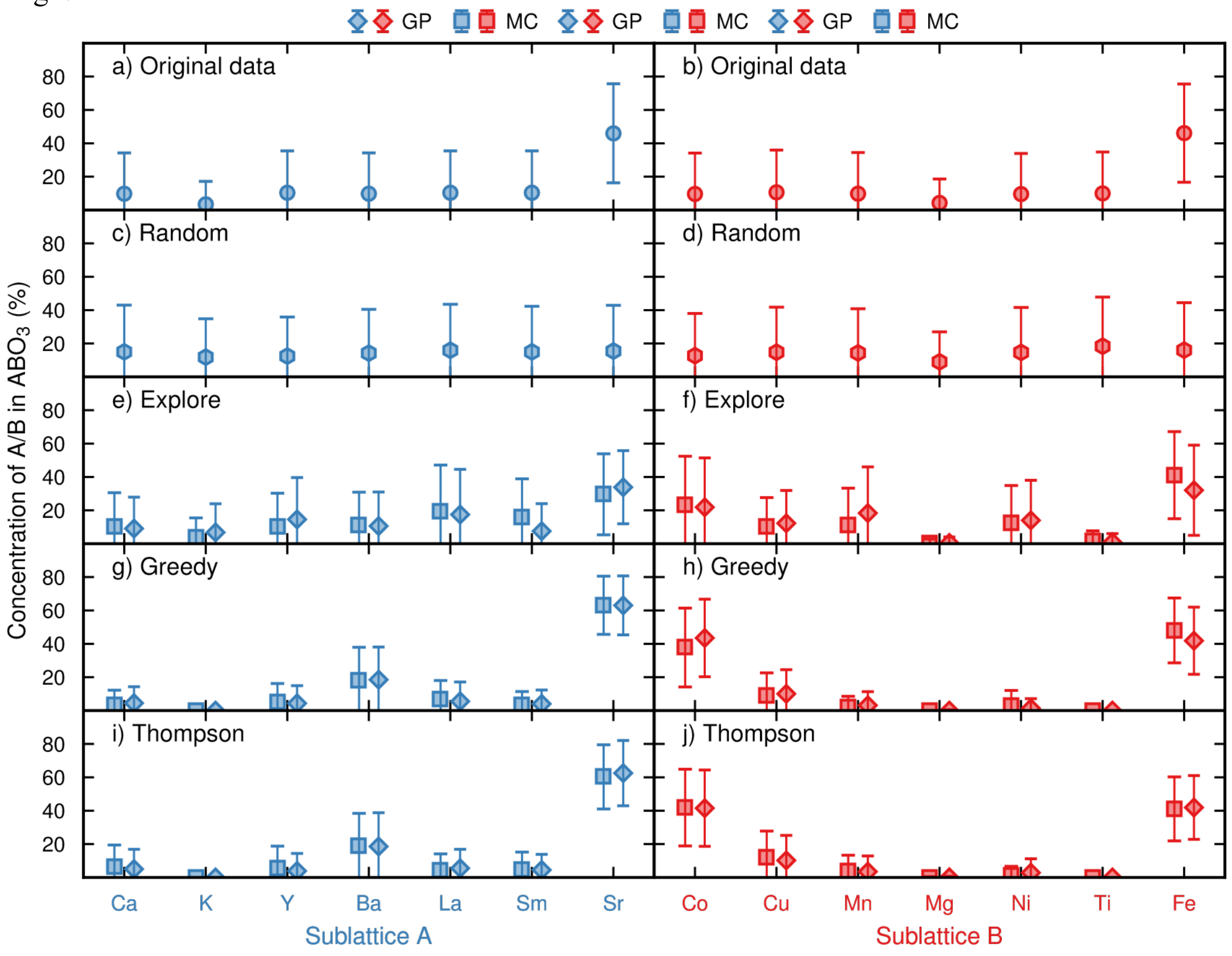

FIG. 9. Diagrams with average counts of elements located on sublattice A (a,c,e,g,i) and B (b,d,f,h,j) in $ABO_3$ OCs, where the bars indicate the corresponding standard deviations, for CLAS. Specifically, this includes results from sampling based on exploration (a-b); a greedy strategy (c-d); and the Thompson method (e-f), combined with either GP (diamonds) or MC dropout (squares), as well as random selection (g-h) and the original dataset (i-j).

### D. Chemical looping oxygen uncoupling

Given the inherent issues related to the use of a probability to assess OC performance as well as the severely restrictions placed on the original training data, it was deemed prudent to go beyond the framework used by Wang et al. [11] and instead consider the problem of finding HEOCs for CLOU. As detailed in section II, the OTC was used as the target property since it governs the ability to transfer oxygen between the air and fuel reactors and, therefore, is a well-defined and common measure of the usability of a given OC. The AL was, moreover, initiated based on a 1.5 times larger database of structures, with random compositions that included as many as seven distinctive metallic elements. These factors were found to have a profound impact on the aggregated results of the AL cycles (see Figure 10). Specifically, the distribution with respect to the OTC shifts to significantly higher values compared to the original data, in terms of both the average and the maximum, when using Thompson or greedy sampling together with GP. As should be expected, the distribution of the configurational entropy is centred around a higher value compared to the original data—which not by design even though the explicit goal is to find HEOCs (see Figure 10b,e). As earlier, the opposite is true for the average abundance (see Figure 10c,f).

A clearer distinction between the two strategies emerges upon analyzing the averages of the properties over the cycles (see Figure 11). In particular, the Thompson method gives a relatively monotonous increase in the OTC, modulated by small fluctuations (see Figure 11b). For a greedy sampling, meanwhile, it is difficult to detect a clear

| Place | Formula | Cycle | Probability | $S_{conf}/R$ | Metal abundance |
|---|---|---|---|---|---|
| 1st | $Co_6Fe_2La_2Sr_6O_{24}$ | 0 | 0.232 | 1.12 | 7.19e+03 |
| 2nd | $Fe_6La_3Mg_2Sr_5O_{24}$ | 0 | 0.231 | 1.22 | 2.41e+04 |
| 3rd | $Fe_5Mg_3Sm_4Sr_4O_{24}$ | 0 | 0.228 | 1.35 | 2.21e+04 |
| 4th | $Ba_4Fe_8Sr_4O_{24}$ | 0 | 0.225 | 0.693 | 2.83e+04 |
| 5th | $FeK_3Mn_7Sr_5O_{24}$ | 0 | 0.225 | 1.04 | 7.97e+03 |
| 6th | $Fe_6La_2Ni_2Sr_6O_{24}$ | 0 | 0.222 | 1.12 | 2.13e+04 |
| 7th | $BaCo_4Fe_4LaSr_6O_{24}$ | 26 | 0.22 | 1.43 | 1.42e+04 |
| 8th | $Ba_2Co_5Fe_3Sr_4Y_2O_{24}$ | 11 | 0.22 | 1.7 | 1.07e+04 |
| 9th | $Ba_2Co_4Fe_4Sr_4Y_2O_{24}$ | 21 | 0.219 | 1.73 | 1.42e+04 |
| 10th | $BaCo_5Fe_3La_2Sr_5O_{24}$ | 19 | 0.218 | 1.56 | 1.07e+04 |
| 11th | $Ba_8Fe_7MgO_{24}$ | 0 | 0.217 | 0.377 | 2.63e+04 |
| 12th | $Cu_6Fe_2Sr_3Y_5O_{24}$ | 0 | 0.217 | 1.22 | 7.14e+03 |
| 13th | $Fe_2Mg_6SrY_7O_{24}$ | 0 | 0.216 | 0.939 | 1.58e+04 |
| 14th | $Fe_4K_2Mn_4Sr_6O_{24}$ | 0 | 0.216 | 1.26 | 1.71e+04 |
| 15th | $Ca_2Co_2Fe_3Mn_3Sr_6O_{24}$ | 19 | 0.214 | 1.64 | 1.61e+04 |
| 16th | $Co_5Fe_3La_2Sr_6O_{24}$ | 25 | 0.213 | 1.22 | 1.07e+04 |
| 17th | $Co_8La_2Sr_6O_{24}$ | 0 | 0.213 | 0.562 | 156 |
| 18th | $Fe_6Ni_2Sr_5Y_3O_{24}$ | 0 | 0.212 | 1.22 | 2.12e+04 |
| 19th | $Ba_5Co_2Fe_6Sr_3O_{24}$ | 2 | 0.212 | 1.22 | 2.13e+04 |
| 20th | $Ca_6Fe_6Sr_2Ti_2O_{24}$ | 0 | 0.211 | 1.12 | 3.74e+04 |

TABLE III. Leaderboard with the top 20 OCs, found via Thompson sampling based on GP, that have been ranked by the probability that they would be suitable for CLAS. The table also includes data on the, first, cycle where the candidate was identified, the configurational entropy ($S_{conf}$/R), and the average metal abundance in the earth's crust ($g_{metal}\,kg^{-1}_{crust}$).

$_2 g_{OC}^{-1}$)

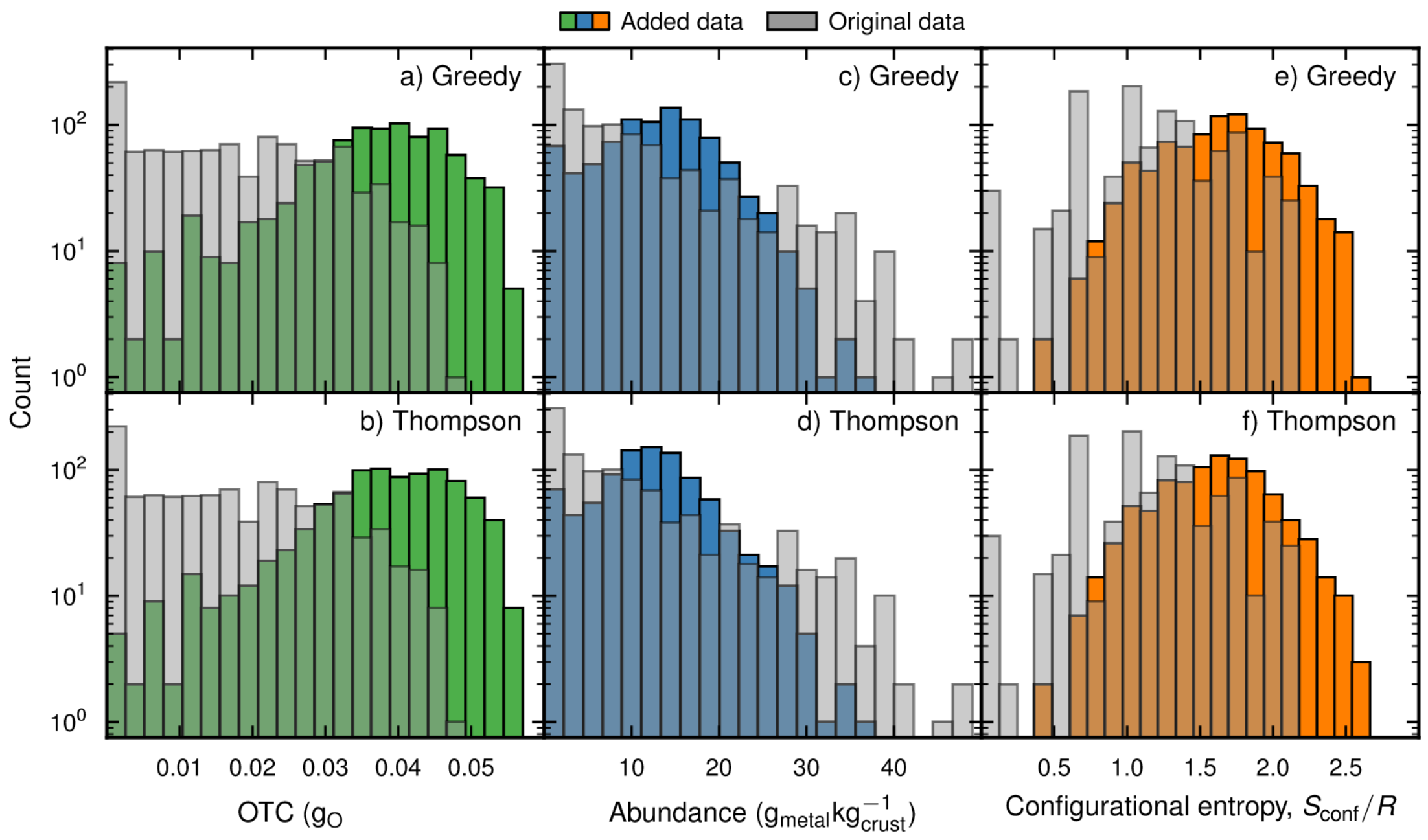

FIG. 10. Histograms based on the OTC (a,d); average abundance of the metallic elements in the earth's crust (b,e); and configurational entropy, $S_{conf}/R$, (c,f) of HEOCs, in which the original dataset is also shown (light grey). Note that the data from the first 10 cycles were compiled from 10 individual runs, after removing duplicate entries.

trend in the average OTC, after an initial, sharp rise; in fact, there are substantial dips at irregular intervals, especially when few new candidates are added, and even a decreasing trend during the last 10 cycles. Thompson sampling would, in other words, have been superior if fewer cycles had been performed. Notwithstanding the cases discussed in the two previous subsections, there is no persistent increase in the configurational entropy, beyond the first few cycles (see Figure 11b-c). Similarly, the average metal abundance remains relatively constant, and even has a tendency to rise slightly rather than declining (see Figure 11e-f). Another interesting observation is that the variations in mean values for all three properties are less pronounced during the first 10 cycles, in spite of the large spread among the individual points, due to the fact that the corresponding data were generated in 10 separate runs. Interestingly, this is followed by a substantial rise in the OTC, especially when using the Thompson method. As such, it could be argued that parallel model training based on random seeds can be efficient initially or when using greedy sampling as this leads to a more stable progression. Even so, it should be stressed that the associated computational demand is substantially higher since many more candidates have to be validated in each cycle. Evidently, this extra effort is associated with a modest learning rate, indicating the need for a sober attitude towards the training of multiple models, in parallel.

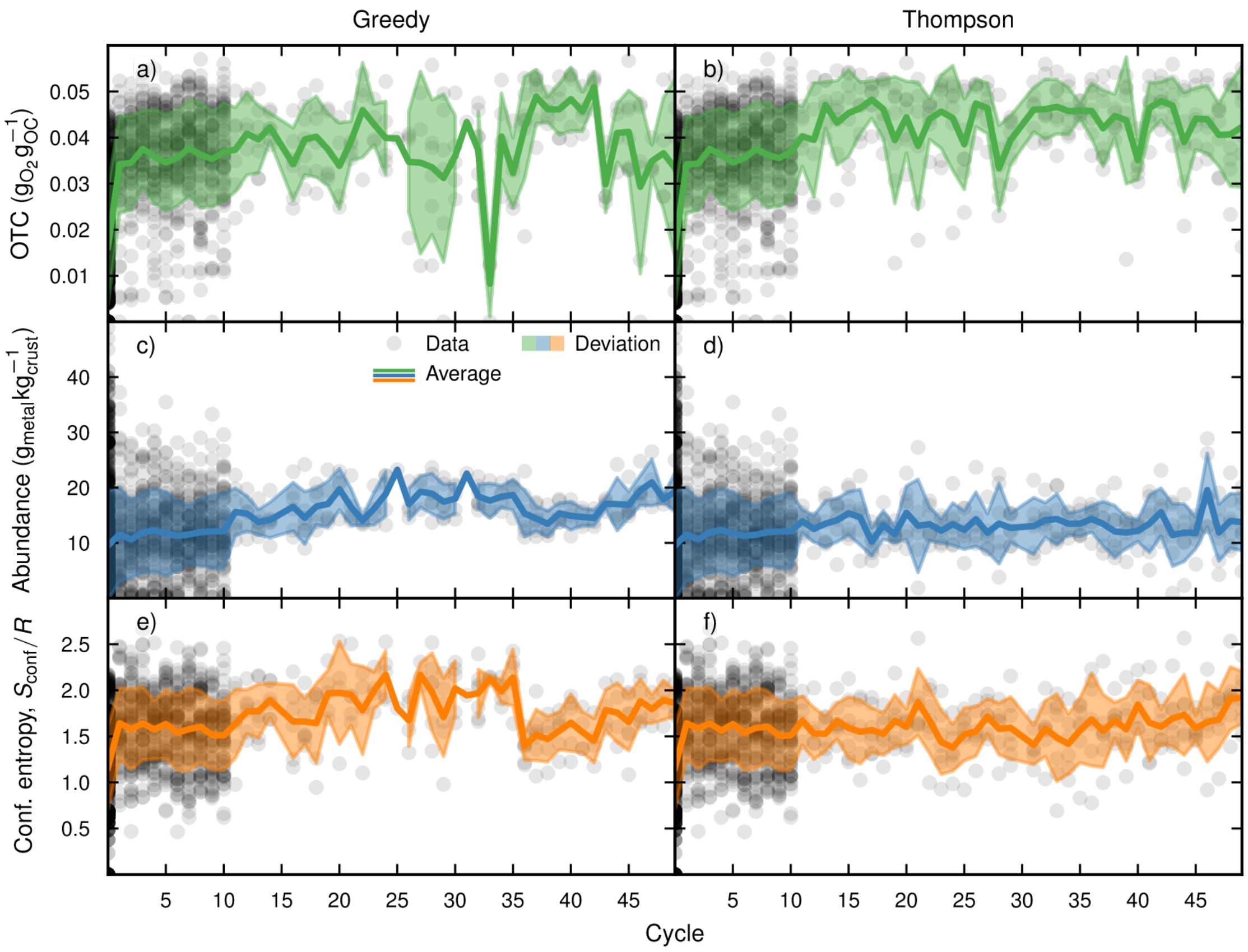

FIG. 11. Scatter plots with the OTC (a-b); average abundance of the metallic elements in the earth's crust (c-d); and configurational entropy, $S_{conf}$/R, (e-f) of the HEOCs in the original dataset (light grey) as well as those added in cycles 1–10, which includes data from 10 individual runs, (purple) and 11–50 (pink). The corresponding averages (solid line) and standard deviations (filled curve) are also shown.

When considering the elemental distribution among the candidate HEOCs, it is crucial to remember that the original training data set consists of completely random compositions, in contrast to the cases discussed in the two previous sections (see Figure 12e-f). The average concentrations should therefore provide a direct measure of which atomic species are the most beneficial for CLOU. Specifically, it is found that Ca, Y, and Mn are by far the most common, both when using a greedy and Thompson-based sampling (see Figure 12a-d). This should be regarded as a validation of both the AL strategy as well as the MLIP-based computations since $CaMnO_3$, and doped variants thereof, have been extensively tested in studies of CLOU [41–45]. The fact that the standard deviations are considerably larger during the first 10 cycles is not unexpected, given that parallelised learning, as mentioned above, leads to a wider spread in terms of the key properties. The candidates from the later cycles should, nonetheless, be regarded as the most interesting since their OTC tends to be higher. It is, hence, noteworthy that significant shifts can be observed when comparing the average compositions from cycles 1-10 and 11-50, respectively. In particular, the concentrations of elements other than Ca, Y, and Mn drop significantly, albeit from low levels. The only exception is Ti, which average actually increases, indicating that this species is a possible dopant, in agreement with reported experiments [42, 45].

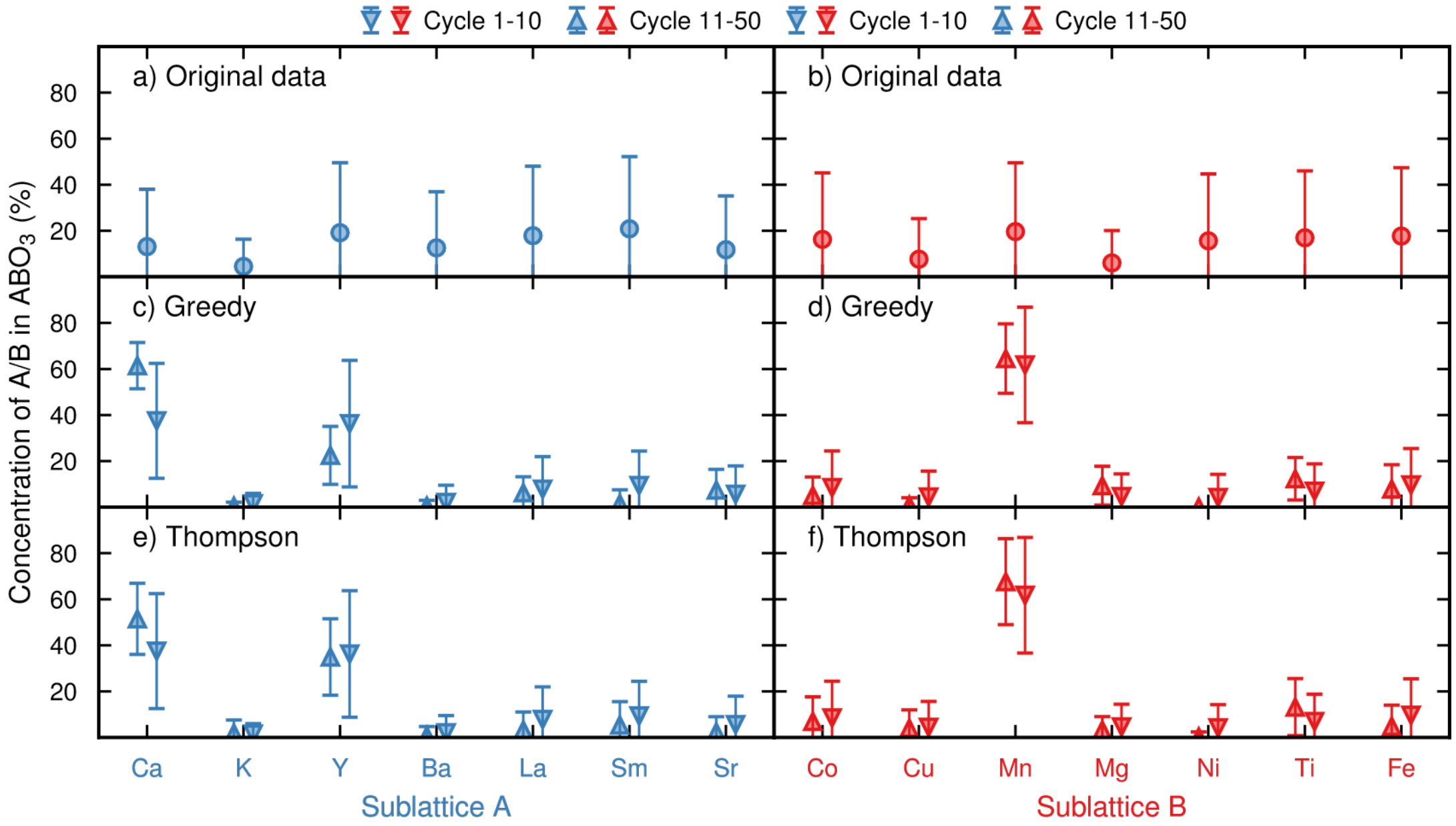

FIG. 12. Diagrams with average counts of elements located on sublattice A (a,c,e) and B (b,d,f) in $ABO_3$ HEOCs, where the bars indicate the corresponding standard deviations. Specifically, this includes data from the original dataset (e-f) as well as samples produced during cycles 1-10 (downward triangles) and 11-50 (upward triangles) using either a greedy strategy (a-b) or the Thompson method (c-d) combined with GP.

Before continuing, it is helpful to restate that the OTC stems directly from the difference in oxygen vacancy concentration between the air and fuel reactors, weighted by the molar mass of the perovskite. As the non-stoichiometry ($\delta$) was only allowed to vary between 0.0 and 0.5, the same holds true for the difference ($\Delta\delta$), which means that the HEOCs can be roughly categorized as having a low, medium or high oxygen release capability if $0 \leq \Delta\delta \leq 0.1875$, $0.1875 \leq \Delta\delta \leq 0.375$ and $0.375 \leq \Delta\delta \leq 0.5$, respectively. By analysing the candidate counts calculated based on these groupings, valuable insights can be gained regarding which atomic species—when added to $Ca_{1-x}Y_xMnO_3$ for the purpose of forming HEOs—allow a high OTC to be maintained (see Figure 13). Although one would anticipate that many candidates containing Ca, Y, and Mn should belong to the “high” category, given the data presented earlier, significant variations can still be observed. For instance, the corresponding count continuously increases with the amount of Mn and reaches its highest value at when this element is if found on all 16 B sites while Ca peaks at nine atoms per $A_{16}B_{16}O_{40}$ supercell and Y at six. Of those that remain, only Sm; Cu; and Ti display a maximum above the lowest limit, more precisely at three; two; and four atoms per supercell, respectively. While a successive decline is observed for the other species, more than a few of the best candidates do contain a limited concentration of La, Sr, Co, Mg, and Fe; in fact, there are quite many ”high“ examples in which either of these, except Sr and Mg, occupy all available sites.

The previous analysis suggests that the most promising materials are based on $Ca_{1-x}Y_xMnO_3$, with $x < 0.5$, in which a few Mn atoms have been replaced by Ti; Co; Fe; Mg; or Cu and some Ca/Y by Sm; La; or Sr. This is perfectly reflected by the 20 highest ranking materials, identified via Thompson sampling based on GP. In fact, this list is almost exclusively made up of $(Ca_{16-y}Y_y)(Mn_{16-x}Ti_x)O_{48}$, where $3 \leq y \leq 7$ and $1 \leq x \leq 6$, with small additions of K as well as Co, Cu, Mg, and Fe; i.e., no more than three and four substitutions on the A and B site, respectively (see Table IV). Moreover, the vast majority have a configurational entropy above the 1.5R threshold and should, thus, be regarded as HEOCs, including the top four candidates: $Ca_{13}CoMgMn_9Ti_5Y_3O_{48}$, $Ca_{11}Mg_3Mn_8Ti_5Y_5O_{48}$, $Ca_{11}FeMg_3Mn_8Ti_4Y_5O_{48}$, and $Ca_{13}Co_2FeMn_9Ti_4Y_3O_{48}$.

## IV. CONCLUSIONS

The study at hand study has not only demonstrated the capability of AL to consistently identify promising materials based on specific conditions. Convincingly, new candidates with equal or enhanced performance were found when a 10% subset was used as a starting point, in spite of the artificial measure, in the form of the probability that a given OC is useful for CLAS or CLDR, and non-optimal set of initial training structures. Especially impressive results were observed in the latter, and presumably more difficult, case. Even so, it is assumed that a similar level of improvement could have been obtained when considering CLAS if more stringent requirements could had been implemented. The fact that a greedy strategy fares better than Thompson sampling, which balances exploration and exploitation, is deemed the direct consequence of either a lucky coincidence or, more probably, prior knowledge: the original training data already covers the limited portion of configurational space with the best materials. It would seem that conventional methods can rival or even excel their more advanced counterparts under such circumstances, namely when the area of interest can be limited based on *a priori* information. Indeed, the initial grid search performed by Wang et al. [11] was evidently sufficient, even though said authors trained a ML model using the resulting data to facilitate the sampling of compositions with additional components.

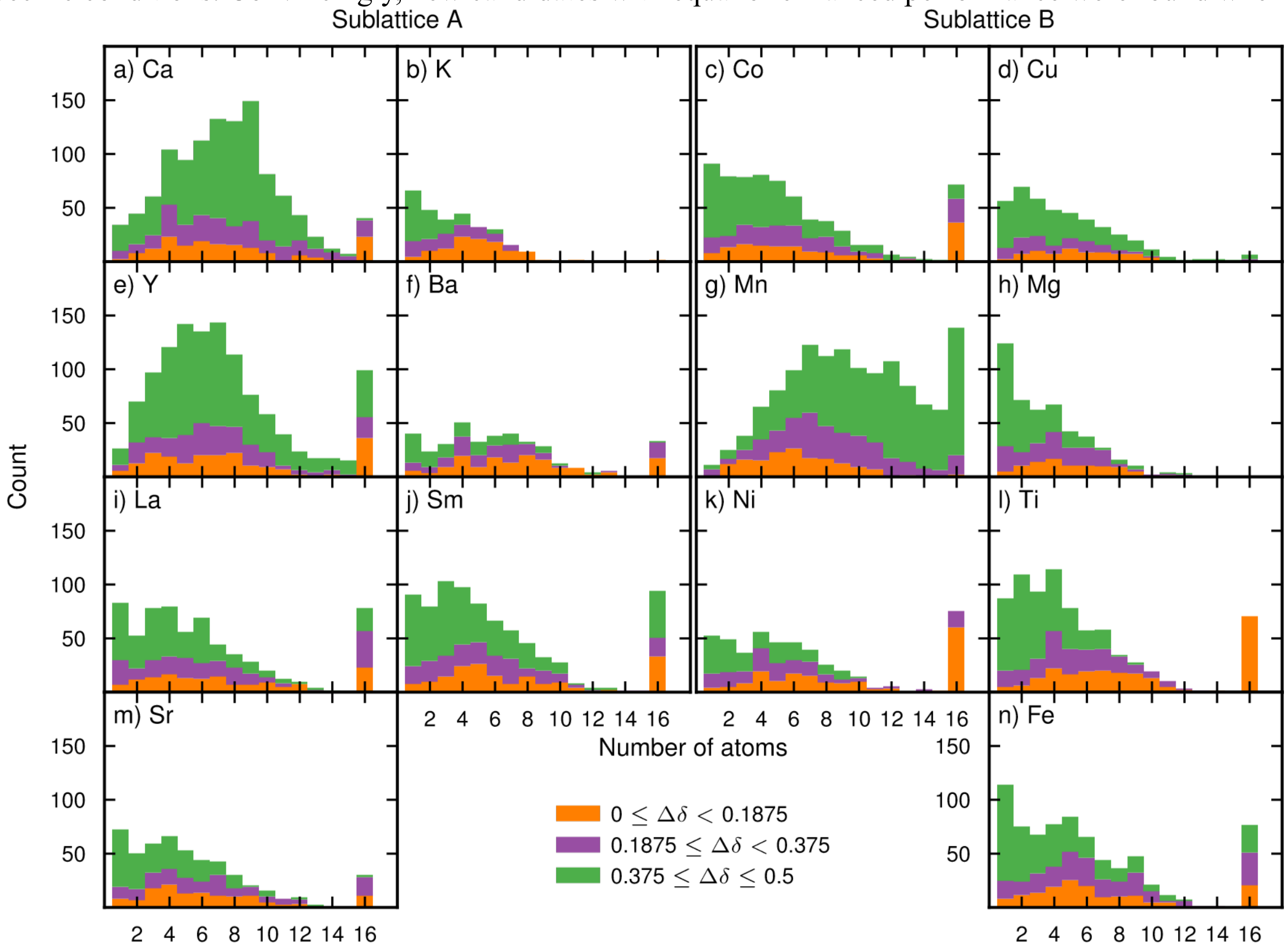

FIG. 13. Bar plots with counts of the number of HEOCs candidates that contain a certain number of Ca (a), K (b), Co (c), Cu (d), Y (e), Ba (f), Mn (g), Mg (h), La (i), Sm (j), Ni (k), Ti (l), Sr (m), and Fe (n) atoms, per $A_{16}B_{16}O_{48}$ supercell. The percentage corresponding to a difference in the non-stoichiometry, between the air and fuel reactors, of $0 \leq \Delta\delta < 0.1875$ (orange), $0.1875 \leq \Delta\delta < 0.375$ (purple), and $0.375 \leq \Delta\delta < 0.5$ (green) have also been indicated.

While the results presented here indicate that the relative efficacy of different AL strategies is strongly correlated with the exact definition and scope of the problem, the ability to balance exploration and exploitation is beneficial, especially in terms of robustness. As the search for CLDR OCs showed, UEs based on GP, as opposed to MC dropout, provide a higher learning rate and average value on the target property. In addition, greedy sampling has a tendency to become trapped and consequently find fewer promising candidates compared to the Thompson method. Even though this issue can be circumvented by training multiple models in parallel, this is accomplished at the cost of a slower progress and larger computational demand.

The insights gained from the initial trial studies were applied to the problem of finding HEOCs for CLOU, which proved to be a very successful enterprise. More precisely, a plentiful number of candidates with better performance and larger configurational entropy compared to the, randomly generated, initial training data were identified, thereby showcasing the true potential of AL-based strategies. In agreement with available experimental data, $CaMnO_3$based materials displayed the highest OTC, especially when doped by species such as Y, Sm, Ti, Fe, and Mg. Approaches such as those used in this study, consequently, serves a plethora of purposes simultaneously: offering a fully automatic framework for identifying specific, promising candidates and, concomitantly, generating valuable qualitative information regarding favorable chemistries. In other words, they can be useful for guiding experimental work either as inline tools, and possibly incorporated into the workflow of completely autonomous laboratories, or to perform, e.g., a pre-study. Regardless of the specific implementation, it is envisioned that the introduction of AL represents a paradigm shift within materials science, with the prospect of boosting the development of green technologies and thereby aiding the transition to a more sustainable society.

| Place | Formula | Cycle | OTC | $S_{conf}/R$ | Metal abundance |
|---|---|---|---|---|---|
| 1st | $Ca_{13}CoMgMn_9Ti_5Y_3O_{48}$ | 8 | 0.057 | 1.52 | 1.87e+04 |
| 2nd | $Ca_{11}Mg_3Mn_8Ti_5Y_5O_{48}$ | 20 | 0.0562 | 1.65 | 1.76e+04 |
| 3rd | $Ca_{11}FeMg_3Mn_8Ti_4Y_5O_{48}$ | 10 | 0.056 | 1.8 | 1.92e+04 |
| 4th | $Ca_{13}Co_2FeMn_9Ti_4Y_3O_{48}$ | 32 | 0.056 | 1.59 | 1.96e+04 |
| 5th | $Ca_{11}CoMg_3Mn_8Ti_4Y_5O_{48}$ | 43 | 0.0559 | 1.8 | 1.74e+04 |
| 6th | $Ca_{11}Co_2Fe_2Mg_2Mn_6Ti_4Y_5O_{48}$ | 49 | 0.0551 | 2.12 | 2.01e+04 |
| 7th | $Ca_{12}Co_4Mn_6Ti_6Y_4O_{48}$ | 4 | 0.055 | 1.64 | 1.68e+04 |
| 8th | $Ca_{12}Co_2Cu_2Mn_7Ti_5Y_4O_{48}$ | 14 | 0.0546 | 1.81 | 1.67e+04 |
| 9th | $Ca_{11}CoCuMgMn_9Ti_4Y_5O_{48}$ | 33 | 0.0543 | 1.81 | 1.6e+04 |
| 10th | $Ca_{11}MgMn_{13}Ti_2Y_5O_{48}$ | 15 | 0.0542 | 1.22 | 1.57e+04 |
| 11th | $Ca_{11}Fe_3Mn_7Ti_6Y_5O_{48}$ | 38 | 0.0541 | 1.66 | 2.08e+04 |
| 12th | $Ca_8K_3Mn_{12}Ti_4Y_5O_{48}$ | 3 | 0.0539 | 1.59 | 1.34e+04 |
| 13th | $Ca_{10}Fe_3Mg_2Mn_8Ti_3Y_6O_{48}$ | 15 | 0.0539 | 1.9 | 2.05e+04 |
| 14th | $Ca_{11}Mn_{12}Ti_4Y_5O_{48}$ | 10 | 0.0539 | 1.18 | 1.53e+04 |
| 15th | $Ca_{11}Co_2Mn_9Ti_5Y_5O_{48}$ | 32 | 0.0538 | 1.57 | 1.54e+04 |
| 16th | $Ca_{11}Fe_4Mn_9Ti_3Y_5O_{48}$ | 16 | 0.0536 | 1.61 | 2.21e+04 |
| 17th | $Ca_9Co_2Fe_2Mg_3Mn_5Ti_4Y_7O_{48}$ | 46 | 0.0535 | 2.23 | 1.82e+04 |
| 18th | $Ca_{13}Mn_{11}Sm_2SrTi_5O_{48}$ | 8 | 0.0535 | 1.22 | 1.81e+04 |
| 19th | $Ca_9Mg_3Mn_{11}Ti_2Y_7O_{48}$ | 37 | 0.0534 | 1.52 | 1.45e+04 |
| 20th | $Ca_{10}KMn_{15}TiY_5O_{48}$ | 39 | 0.0534 | 1.06 | 1.42e+04 |

TABLE IV. Leaderboard with the top 20 HEOCs, found via Thompson sampling based on GP, that have been ranked by the OTC. The table also includes data on the, first, cycle where the candidate was identified, the configurational entropy ($S_{conf}$/R), and the average metal abundance in the earth's crust ($g_{metal}\ kg^{-1}_{crust}$).

**ACKNOWLEDGMENTS**

This work was funded by the Swedish Research Council (2020-03487) and the computations were enabled by resources provided by the Swedish National Infrastructure for Computing (SNIC) at NSC and C3SE partially funded by the Swedish Research Council through grant agreement no. SNIC 2021/3-41, SNIC 2021/5-623, SNIC 2021/5-561, and SNIC 2022/5-156.

---